\documentclass{article} %
\usepackage{iclr2027_conference,times}

\usepackage{amsmath,amsfonts,bm}

\def\eqref#1{equation~\ref{#1}}

\def\1{\bm{1}}

\DeclareMathAlphabet{\mathsfit}{\encodingdefault}{\sfdefault}{m}{sl}
\SetMathAlphabet{\mathsfit}{bold}{\encodingdefault}{\sfdefault}{bx}{n}

\usepackage{hyperref}
\hypersetup{pdftitle={RankCast: Phase-Aware Rank Balancing in MoE Serving},
  pdfauthor={Kaidi Fu, Xu-Hao Lu, Pingwei Sun, Jianchao Tan, Yuchen Xie, Xunliang Cai}}
\usepackage{url}
\usepackage{graphicx}
\usepackage{amsmath}
\usepackage{amssymb}
\usepackage{booktabs}
\usepackage{enumitem}
\usepackage{needspace}

\title{RankCast: Phase-Aware Rank Balancing in MoE Serving}

\author{
Kaidi Fu$^{1}$, Xu-Hao Lu$^{2}$, Pingwei Sun$^{2}$, Jianchao Tan$^{2,*}$, Yuchen Xie$^{2}$, Xunliang Cai$^{2}$ \\
$^{1}$University of Illinois at Urbana-Champaign \qquad $^{2}$Meituan \\
\texttt{kaidifu2@illinois.edu} \qquad \texttt{tanjianchao02@meituan.com}
}

\iclrfinalcopy
\begin{document}

\maketitle
\lhead{}
\renewcommand{\headrulewidth}{0pt}
{\renewcommand{\thefootnote}{\fnsymbol{footnote}}\footnotetext[1]{Corresponding author.}}

\begin{abstract}
Modern large language model (LLM) inference increasingly adopts disaggregated architectures that
separate Prefill (P) and Decode (D) phases. Meanwhile, Expert Parallelism (EP) has been widely
adopted in large Mixture-of-Experts (MoE) model serving. However, load imbalance still slows down
MoE inference because each layer must wait for the busiest EP rank. The main
source of latency differs between prefill and decode. Prefill is compute-bound, and its MoE layers'
latency is determined by the maximum token count of the busiest rank. At the same time, under moderate batch sizes,
Decode is memory-bound, so the latency is dominated by the maximum active-expert count across ranks. To
solve both problems under a unified framework, we present RankCast, a system that reduces
rank-level maximum token counts during prefill and active-expert counts during decode. For every
token, RankCast preserves the predicted Top-$p$ experts and chooses the remaining $k-p$ from its
Top-$L$ candidates ($L>k$). In prefill, RankCast yields a price per rank that steers flexible
assignments to less loaded ranks. In decode, RankCast closes experts on the busiest ranks while
preserving the constraints for each token. Both paths are implemented and tested in the
state-of-the-art inference frameworks SGLang~\citep{zheng2024sglang} and DeepEP~\citep{deepep2025}. We overlap
RankCast's kernels with model execution and reuse the existing DeepEP communication channel for
metadata transmission. On Qwen3-235B-A22B with EP$=8$ and Qwen3.5-397B-A17B with EP$=16$,
RankCast removes over 75\% of the busiest rank's excess prefill load, improves prefill input
throughput by up to 19\%, and reduces decode time per output token by up to 10\%. The quality cost
is below one percentage point.

\end{abstract}

\section{Introduction}
\label{sec:introduction}

Mixture-of-Experts (MoE) architectures are used in model families including Mixtral, LongCat,
DeepSeek-V3/V4, Qwen3, and
Kimi~\citep{jiang2024mixtral,longcat,deepseekai2024deepseekv3,deepseekai2026deepseekv4,qwen3,kimiteam2026k3}. In MoE model
serving, expert parallelism (EP) partitions expert weights across ranks and exchanges token
activations through dispatch and combine collectives. These synchronization barriers cause the
layer's latency to be bottlenecked by the busiest rank.

Although a lot of models deploy per expert imbalance control algorithms including auxiliary loss and
quantile balancing, during serving time,
the frozen router still faces batch-level variation and domain drift, and what really matters here
is the instantaneous load on each physical EP rank rather than the long-horizon utilization of each
expert. DeepSeek-V3 notes that domain shift can cause this imbalance at inference and deploys
redundant experts in serving to counter it~\citep{deepseekai2024deepseekv3}. UltraEP reports
pre-balancing inter-rank imbalance ratios of $1.30$--$4.01$ in large-EP training and serving-prefill
workloads and observes that expert popularity shifts across microbatches, layers, and data
domains~\citep{wei2026ultraep}.

PD disaggregation is also widely adopted in modern MoE model serving. However, Prefill and decode
have different balancing objectives. In prefill, computation and communication grow linearly with
the number of tokens a rank receives, and the layer's total latency is determined by the slowest
rank due to synchronization barriers. In decode at moderate batch sizes, expert GEMM time is
dominated by loading the weights of active experts, and additional tokens matter little until the
tokens assigned to a single expert exceed one block's capacity. Our profiling
(Appendix~\ref{sec:eval_profile}) confirms both regimes. Reducing prefill latency therefore
requires minimizing the maximum token load across ranks, whereas reducing decode latency requires
minimizing the maximum number of active experts across ranks. OEA likewise observes that MoE decode latency scales
approximately linearly with the number of active experts in this
regime~\citep{oncescu2025opportunistic}.

\begin{figure}[t]
    \centering
    \includegraphics[width=\linewidth]{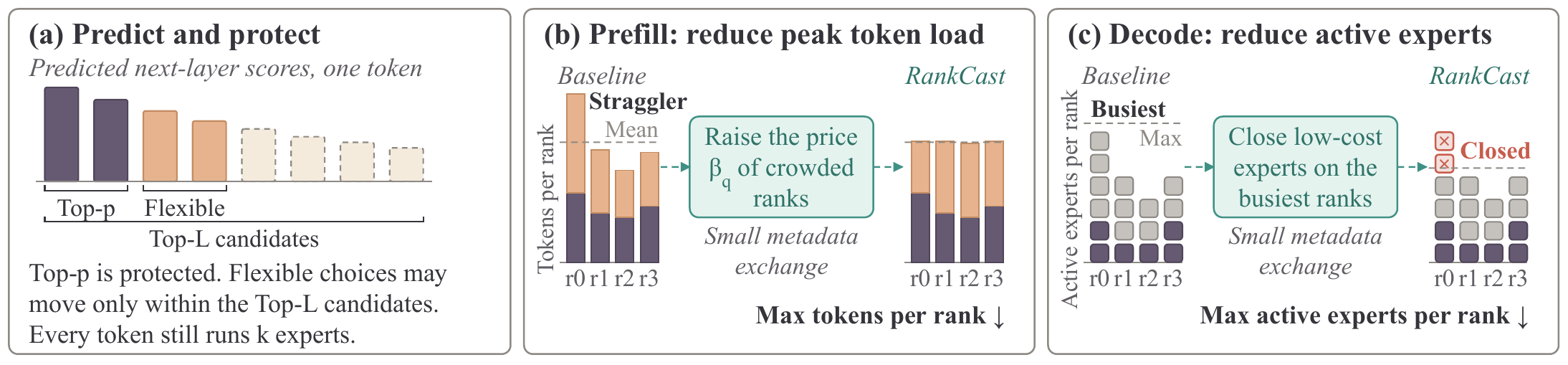}
    \caption{RankCast overview. (a) Predicted next-layer routing defines protected Top-$p$ experts and Top-$L$ candidates. (b) Prefill uses rank prices to redistribute flexible assignments and reduce peak token load. (c) Decode closes low-cost experts to reduce the peak active-expert count while preserving protected experts and at least $k$ candidates per token.}
    \label{fig:rankcast_overview}
\end{figure}

Current inference systems typically address this imbalance challenge through expert placement,
replication, and predictive
prefetching~\citep{deepseekai2025eplb,han2025gracemoe,yang2026libra,zhu2026probe}. These approaches
preserve the router's expert choices but exchange memory or communication bandwidth for load
balance.

UltraEP shows that placement built from historical load cannot track these fast shifts and
can even worsen imbalance~\citep{wei2026ultraep}. Replicas occupy HBM that would otherwise hold KV cache; CRAFT
reports that conventional replication reduces available KV-cache capacity by $19\%$--$75\%$ and can
lower goodput when the concurrency loss exceeds the balancing gain~\citep{zhao2026craft}.
Prefetching avoids permanent residency but must fit expert-weight transfers into a short hiding
window and schedule them around token all-to-all~\citep{zhu2026probe}; across nodes, UltraEP deems
moving expert state over the slower scale-out network impractical on the hot
path~\citep{wei2026ultraep}. Both costs are acute in long-context serving, where
KV-cache capacity is scarce and KV-cache transmission adds additional communication pressure.

Rerouting tokens to experts on less loaded ranks in Prefill or to already active experts in Decode
provide a new way to avoid both costs. This is the key insight behind RankCast. However, changing a
frozen model's expert choices could affect output quality. \textsc{RankCast}
(Figure~\ref{fig:rankcast_overview}) therefore limits rerouting through a protected Top-$p$ prefix and a
Top-$L$ candidate set, with $p\leq k\leq L$, so balancing changes only lower-ranked choices while
keeping $k$ experts per token. For prefill, each rank receives a price that is subtracted from the
scores of the candidates it hosts, so crowded ranks receive fewer token assignments.

Decode requires a different objective because token prices do not directly reduce the cost of
loading expert weights. RankCast starts from the experts activated by the Top-$k$ routing
and closes experts on the busiest ranks (Figure~\ref{fig:rankcast_overview}c), keeping every Top-$p$ expert open and leaving each token at least $k$ candidates from its 
Top-$L$ set. The target router then selects its Top-$k$ using its own scores with closed experts
suppressed.

To make balancing decisions available before routing, both phases predict the next layer's routing
and carry compact metadata through DeepEP's existing dispatch communication, so price and mask
computations overlap with model execution (Section~\ref{sec:system}) and add no synchronization
barrier.

\needspace{4\baselineskip}
\textbf{Contributions.}
\begin{itemize}[leftmargin=*,itemsep=5pt,topsep=5pt]
    \item We characterize and formulate EP balancing as a phase-aware, rank-level problem: prefill is
          bottlenecked by maximum per-rank token load, while decode is dominated by the maximum number of active experts across ranks.
    \item We develop a prefill algorithm that preserves each token's Top-$p$ prefix and balances the remaining
          assignments with compact distributed histograms and rank prices.
    \item We develop a decode algorithm that closes experts to lower the peak active-expert count per rank while
          preserving each token's protected experts and at least $k$ candidates.
    \item We implement both paths in SGLang and DeepEP and evaluate them end-to-end. On Qwen3-235B-A22B with
          EP$=8$ and Qwen3.5-397B-A17B with EP$=16$, RankCast removes over 75\% of the busiest rank's
          excess prefill load, improves prefill input throughput by up to 19\%, and reduces decode TPOT by up to
          10\%, at a quality cost below one percentage point.
\end{itemize}

\section{Related Work}
\label{sec:related-work}

\textbf{Training-time balancing.}
Switch Transformers, BASE, and Expert Choice balance expert utilization through auxiliary losses,
balanced assignment, or expert-side
selection~\citep{fedus2022switch,lewis2021base,zhou2022expertchoice}. Auxiliary-loss-free balancing
instead updates per-expert routing biases from observed loads~\citep{wang2024auxlossfree}, which
DeepSeek-V3/V4 combine with a small sequence-wise
loss~\citep{deepseekai2024deepseekv3,deepseekai2026deepseekv4}. Kimi K3's Quantile Balancing
derives each expert's training-time bias from the dual of a balanced-assignment problem, which
makes it a quantile of router-score margins that K3 estimates from all-reduced
histograms~\citep{kimiteam2026k3}; RankCast's prefill prices play the same dual role for physical
ranks at serving time.

\textbf{Placement, replication, and prefetching.}
Placement methods keep logical routing but change the expert-to-device mapping: DeepSeek-V3 and
EPLB replicate and rearrange hot experts from estimated
loads~\citep{deepseekai2024deepseekv3,deepseekai2025eplb}, MoETuner solves a placement ILP over
compute and communication skew~\citep{go2025moetuner}, Semantic Parallelism co-schedules placement
with rebatching~\citep{li2026semantic}, and Lina allocates more devices to experts predicted to be
popular~\citep{li2023lina}.

Replication and prefetching adapt capacity to hot experts: GRACE-MoE, CRAFT, and Libra allocate
replicas by load skew over affinity-based groups, per-layer benefit, or
lookahead~\citep{han2025gracemoe,zhao2026craft,yang2026libra}; PROBE prefetches predicted hot
experts within a hiding window~\citep{zhu2026probe}, and MoonEP prefetches redundant experts planned
from the current router outputs~\citep{moonshotai2026moonep}; UltraEP gathers exact
post-gating loads per microbatch and jointly plans replication and
rerouting~\citep{wei2026ultraep}. All of them trade replica storage or expert-state movement for
balance.

\textbf{Inference-time rerouting.}
Capacity-aware inference~\citep{he2026capacity} limits the number of tokens assigned to each expert and drops or reroutes 
any tokens that exceed that expert's capacity; LASER reroutes ambiguous tokens 
to less-loaded experts within a score-derived viable set~\citep{shahout2025laser}. 
Both target token-count balance, the prefill
objective. RankCast also reroutes at inference time but balances physical-rank load and changes
only lower-ranked choices within a limited candidate set.

\textbf{Decode-time active-expert consolidation.}
In memory-bound decode, activating another expert can cost more than adding tokens to an active
one. Lynx remaps low-affinity assignments onto a smaller active set~\citep{gupta2024lynx}; SERE
redirects secondary assignments to similar active experts~\citep{wu2026sere}; OEA lets additional
assignments piggyback on experts the batch already activates~\citep{oncescu2025opportunistic}.
XShare is the closest comparison for the EP objective: it seeds a shared pool with each token's
top-ranked experts, grows it by batch-summed gating score up to a preset budget, and, in its
EP-aware variant, applies the budget per GPU~\citep{vankov2026xshare}. RankCast instead searches for closures that reduce the maximum
active-expert count across ranks subject to each token's predicted candidate-set and
protected-prefix constraints. It validates closures jointly because closures on different ranks can
affect the same token, and it overlaps the search with model execution through DeepEP
(Section~\ref{sec:system}).

\section{Phase-Aware RankCast Algorithms}\label{sec:algorithm}

In the previous sections, we discussed the different optimization objectives for Prefill and Decode. For Prefill,
we reduce the maximum token load across ranks and for decode we reduce the maximum active experts of all ranks.
In this section, we develop separate algorithms under the same routing constraint. For token $i$
with router scores $s_i$, the candidate set and the protected set are
\begin{equation}
    \label{eq:routing_constraint}
    \mathcal C_i=\operatorname{Top}_L(s_i),
    \qquad
    \mathcal F_i=\operatorname{Top}_p(s_i),
    \qquad
    p\leq k\leq L,
\end{equation}
and every token's final selection $\mathcal E_i$ satisfies
$\mathcal F_i\subseteq\mathcal E_i\subseteq\mathcal C_i$ with $|\mathcal E_i|=k$. Both phases keep
each token's Top-$p$ experts and draw the rest from its Top-$L$ candidates.

RankCast computes rank prices in prefill and an expert
mask in decode from predicted next-layer scores. The target layer uses its own scores to select the
final experts (Section~\ref{sec:cross_layer_prediction}).

\subsection{Rankcast for Prefill}
\label{sec:prefill_algorithm}

Prefill balances rank load by adjusting each token's $d=k-p$ flexible assignments.
Consider $N$ tokens and $E$ experts distributed across $P$ EP ranks. Let
$\rho(e)$ be the rank hosting expert $e$, and let $\mathcal E_q$ be the set of experts on rank $q$.
Each token requires $d$ flexible assignments. We therefore retain only its top-$d$ flexible candidates on each rank,
as any candidate outside this set cannot appear in the final selection.

For token $i$ and rank $q$, denote their scores by $a_{i,q,1}\geq\cdots\geq a_{i,q,d}$.
Missing slots are padded with $-\infty$ and cannot be selected. Let $c_q$ be rank $q$'s target load
for flexible assignments, with $\sum_q c_q=Nd$. The target $c_q$ is computed from the balance target after excluding the
protected set (Appendix~\ref{app:capacity}). We choose assignments to maximize their
total score while giving each token exactly $d$ choices and matching each rank's target. With
$x_{i,q,j}\in\{0,1\}$ indicating whether a candidate is selected, the problem can be defined as:
\begin{equation}
    \label{eq:prefill_primal}
    \begin{aligned}
        \max_{\mathbf{x}}\quad & \sum_{i,q,j}a_{i,q,j}x_{i,q,j}                                \\
        \mathrm{s.t.}\quad     & \sum_{q,j}x_{i,q,j}=d\ \ \forall i,
        \qquad \sum_{i,j}x_{i,q,j}=c_q\ \ \forall q.
    \end{aligned}
\end{equation}
This transportation problem has an integral LP optimum whenever a feasible assignment exists
(Appendix~\ref{app:price_derivation}). We therefore solve it through its dual. Let $\alpha_i$ be
the threshold of token $i$ and $\beta_q$ be the price of rank $q$. At the optimum, each token
selects its top-$d$ candidates by the price-adjusted score $a_{i,q,j}-\beta_q$. The threshold
$\alpha_i$ separates its $d$-th and $(d+1)$-th adjusted scores.

RankCast computes both variables in a single sweep, starting from zero prices.
Each token first sets $\alpha_i$ between its $d$-th and $(d+1)$-th largest flexible scores.
Let $w_{i,q,j}=a_{i,q,j}-\alpha_i$ denote the residual score, which measures how far a candidate
lies above or below its token's threshold. A candidate on rank $q$ is selected when its residual
score exceeds $\beta_q$. Each rank therefore sets $\beta_q$ between its $c_q$-th and
$(c_q+1)$-th largest residual scores, so that it receives exactly its target load $c_q$ under the
current thresholds.
With $a_{i,(m)}^{\downarrow}$ denoting the $m$-th largest score among token $i$'s flexible slots
and $w_{q,(m)}^{\downarrow}$ the $m$-th largest residual score on rank $q$, the updates can be
written as:
\begin{equation}
    \label{eq:prefill_coordinate_updates}
    \begin{aligned}
        \alpha_i         & = \frac{1}{2}\bigl(a_{i,(d)}^{\downarrow}+a_{i,(d+1)}^{\downarrow}\bigr),      \\
        \beta_q^{\star}  & = \frac{1}{2}\bigl(w_{q,(c_q)}^{\downarrow}+w_{q,(c_q+1)}^{\downarrow}\bigr).
    \end{aligned}
\end{equation}
A rank requested by many tokens therefore receives a high price, which moves part of its load
to other ranks.

The threshold update needs only the scores of token $i$. The price update, however, needs the
residual scores of all tokens on rank $q$, which are spread across source ranks. RankCast
therefore exchanges more compact histograms instead of individual scores. Let $\tau_0<\cdots<\tau_B$
delimit $B$ shared bins, and let $H_{q,b}$ be the number of local residual scores $w_{i,q,j}$ on rank
$q$ that fall in bin $[\tau_b,\tau_{b+1})$. Each source rank also reports its protected loads
$f_q=\sum_i|\mathcal F_i\cap\mathcal E_q|$, for every rank to compute the same targets
$c_q$. The exchanged data size depends only on $P$ and $B$ (Section~\ref{sec:prefill_implementation}).

After summing the histograms, we scan the bins of rank $q$ from highest to lowest until the count
reaches $c_q$ and interpolate within the last bin. This gives an estimate $Q_q(c_q)$ of
$w_{q,(c_q)}^{\downarrow}$ with error at most one bin width (Appendix~\ref{app:histogram_error}).
Since a common shift of all prices does not change any token's ranking, the final price is
centered across ranks:
\begin{equation}
    \label{eq:prefill_price}
    \beta_q=Q_q(c_q)-\frac{1}{P}\sum_{q'=1}^{P}Q_{q'}(c_{q'}).
\end{equation}
Boundary cases, where $c_q=0$ or all candidates on rank $q$ must be selected, set its price to
the upper or lower end of the histogram range (Appendix~\ref{app:histogram_error}).

At the target layer, the router keeps the protected experts and selects the remaining $d$ experts by their price-adjusted scores:
\begin{equation}
    \label{eq:prefill_selection}
    \mathcal{E}^{\mathrm{prefill}}_i
    =
    \mathcal{F}_i
    \cup
    \operatorname{Top}_{d}
    \{s_{i,e}-\beta_{\rho(e)}\mid e\in\mathcal{C}_i\setminus\mathcal{F}_i\}.
\end{equation}

\subsection{Rankcast for Decode}
\label{sec:decode_algorithm}

For decode, RankCast starts with the experts selected by Top-$k$ routing, and closes experts to reduce the
active-expert count on the busiest ranks. A closed expert cannot be selected by any token.
Each token is subject to the same constraints as in prefill. Let
$\mathcal A_0=\bigcup_i\operatorname{Top}_k(s_i)$ be the initial expert set and $\mathcal
    P=\bigcup_i\mathcal F_i$ the protected set. The objective is
\begin{equation}
    \label{eq:decode_active_set_problem}
    \begin{aligned}
        \min_{\mathcal{A}\subseteq\mathcal{A}_0}\quad & M(\mathcal{A})=\max_r |\mathcal{A}\cap\mathcal{E}_r| \\
        \mathrm{s.t.}\quad                            & \mathcal{P}\subseteq\mathcal{A},
        \qquad
        |\mathcal{C}_i\cap\mathcal{A}|\geq k \quad \forall i.
    \end{aligned}
\end{equation}

RankCast solves this problem greedily. Let $\sigma_i=|\mathcal C_i\cap\mathcal A|-k$ be the slack
of token $i$, which counts the candidates it can still lose. An active, unprotected expert $e$ is
\emph{closable} if $\sigma_i>0$ for every token $i$ with $e\in\mathcal C_i$. Let the quality weight
$\omega_{i,e}\in[0,1]$ denote token $i$'s normalized routing-weight loss estimate for removing
expert $e$. Each rank orders its closable experts by
increasing closure cost:
\begin{equation}
    \label{eq:decode_close_cost}
    \operatorname{cost}(e)
    =
    \sum_{\substack{i\\e\in\mathcal{C}_i}}
    \frac{\omega_{i,e}}{\sigma_i+1}.
\end{equation}
The cost is higher for candidates with larger weights and tokens with fewer alternatives.

Let $A_r=|\mathcal A\cap\mathcal E_r|$ be the number of active experts on rank $r$ and
$M=\max_rA_r$ the count on the busiest rank, which Equation~\ref{eq:decode_active_set_problem}
minimizes. One rank cannot lower $M$ alone when another rank has the same count, so each round
closes experts on several ranks together. Let $h_r=M-A_r$ be the headroom of rank $r$ below the
busiest rank. To lower $M$ by $\Delta$, every rank must keep at most $M-\Delta$ active experts, so
rank $r$ closes its $\Delta-h_r$ cheapest closable experts, or none if $\Delta\leq h_r$. Let
$e_{r,j}$ be the $j$-th cheapest closable expert on rank $r$. Rank $r$ therefore closes $e_{r,j}$
once $\Delta\geq h_r+j$. We call $\operatorname{lev}(e_{r,j})=h_r+j$ the level of $e_{r,j}$, and set
$\operatorname{lev}(e)=+\infty$ for non-closable experts.

The maximum $\Delta$ can be derived by considering two bounding conditions.
First, rank $r$ has only $m_r$ closable experts,
therefore $\Delta\leq h_r+m_r$. Second, experts closed together must not exhaust a token's slack. Token
$i$ can lose at most $\sigma_i$ candidates, so $\Delta$ must stay below $\Lambda_i$, the
$(\sigma_i+1)$-th smallest level among its candidates. The largest feasible reduction is therefore
\begin{equation}
    \label{eq:decode_prefix_delta}
    \Delta^\star
    =\max\Bigl\{0,\;\min\bigl(\min_r(h_r+m_r),\;\min_i(\Lambda_i-1)\bigr)\Bigr\}.
\end{equation}

When $\Delta^\star=0$, a repair step tries other experts on the busiest ranks
(Appendix~\ref{app:decode_frontier_repair}). The solver repeats these steps until $M$ stops
decreasing or the total reduction reaches an optional depth cap $D$. The closed experts form the
router mask of the target layer.

\section{System Implementation}\label{sec:system}

We implement RankCast in SGLang and integrate it with DeepEP's normal prefill path and low-latency
decode path. The implementation must meet three requirements: the balancing decision must be ready
before the next router runs, it must add no collective or synchronization barrier to the
per-layer critical path, and every rank must reach the same balancing decision.

\begin{figure}[t]
    \centering
    \includegraphics[width=\linewidth]{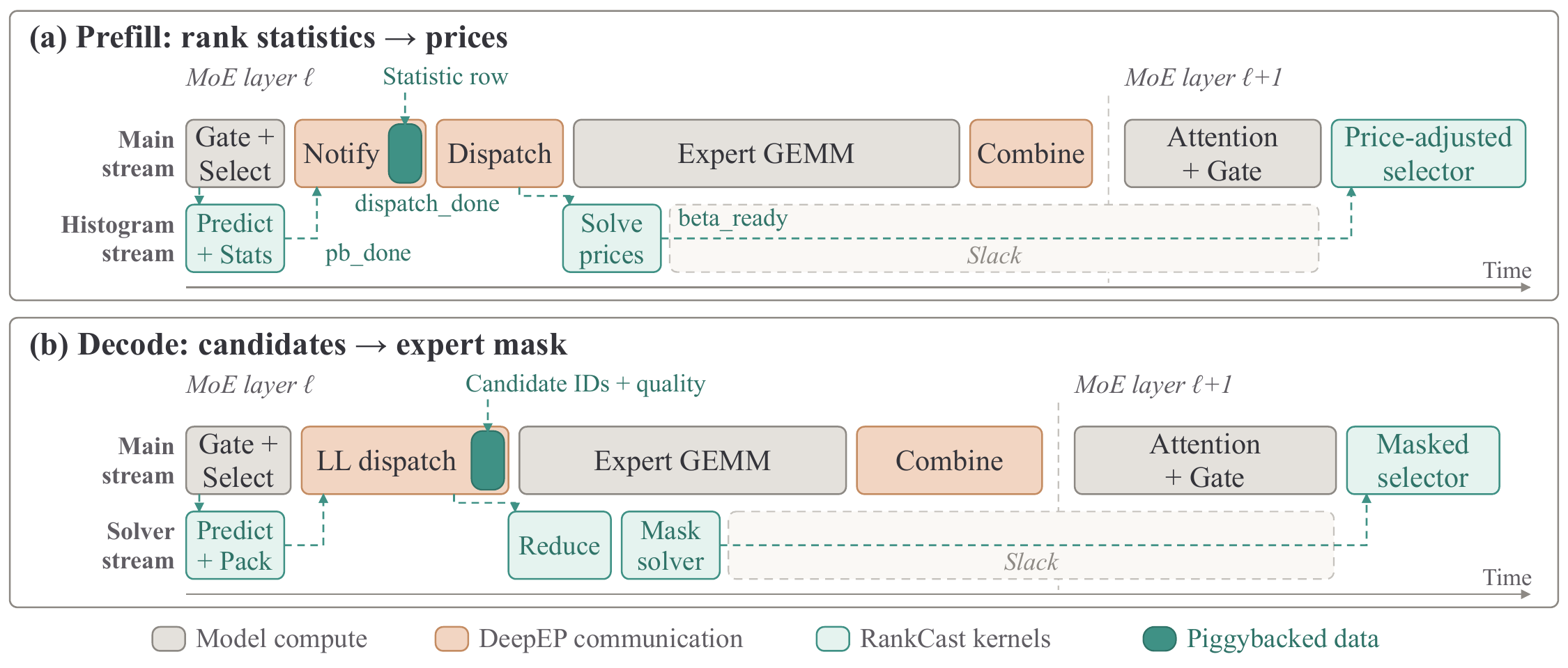}
    \caption{Cross-layer execution of RankCast on one EP rank. Prediction and balancing overlap with model execution, using metadata carried by existing DeepEP communication to compute (a) prefill rank prices or (b) a decode expert mask for the next layer. Dashed dependencies indicate CUDA events; \emph{slack} denotes the interval between solver completion and next-layer selection. LL: low-latency. Timelines are schematic.}
    \label{fig:rankcast_system}
\end{figure}

Both phases therefore use the
four-stage pipeline in Figure~\ref{fig:rankcast_system}. They predict the next layer's routing
scores, exchange compact metadata, compute the balancing decision, and apply it at the next router.
Prefill exchanges rank statistics to compute prices, and decode exchanges candidate lists to
compute an active-expert mask. Both kinds of metadata ride on the token-count messages that
DeepEP already exchanges. Every rank receives the same data and runs the same
deterministic computation, so prices and masks need no separate broadcast.

\suppressfloats[t]

\paragraph{Shared cross-layer router prediction.}\label{sec:cross_layer_prediction}
The balancing decision must be ready before the target layer $\ell+1$ routes its tokens, so
RankCast predicts this routing at the current layer $\ell$. As soon as the current hidden states
$\mathbf X^{(\ell)}$ are available, a side stream applies the next MoE layer's gate weights,
\begin{equation}
    \label{eq:cross_layer_prediction}
    \widehat{\mathbf S}^{(\ell+1)}=\mathbf X^{(\ell)}\mathbf W_{\mathrm{gate}}^{(\ell+1)\top}+\mathbf 1\mathbf b^{(\ell+1)\top},
\end{equation}
and then the solvers run on these predicted scores while the main stream continues.

The prediction determines only the balancing decision. In prefill the decision is a per-rank price.
The target layer takes the protected set and the candidate set from its own scores
(Equation~\ref{eq:prefill_selection}), so a prediction error changes only the price and every token
keeps its true Top-$p$ exactly. In decode the decision is an expert mask, fixed before the target
layer runs, so an error can close a protected expert. Exact top-2 agreement between predicted and
measured scores is 48--66\%, but the true top-2 lies in the predicted Top-$L$ for 98.0--99.8\% of
tokens. A mis-ranked pair usually swaps experts of similar weight, so 87.9--97.8\% of tokens keep
both true top-2 experts after shedding (Table~\ref{tab:prediction}).
Three safeguards bound a wrong prediction. A price in an edge bin of the histogram leaves that
rank's previous price in place, the solver keeps at least $k$ candidates open per token, and a row
with fewer than $k$ valid candidates keeps all of that token's experts open.

RankCast adds
two online corrections (Appendix~\ref{app:online_corrections}). The bias $\mathbf b^{(\ell+1)}$ tracks the mean gap between
predicted and measured router scores, and in prefill a per-rank offset raises the price of ranks
that stay overloaded. Each is updated every few forwards with one all-reduce, outside CUDA graphs
and off the per-layer critical path.

\paragraph{Histogram and price pipeline for prefill.}\label{sec:prefill_implementation}
Prefill runs the four stages of Figure~\ref{fig:rankcast_system}(a) on a side stream. First, a
statistics kernel evaluates Equation~\ref{eq:cross_layer_prediction} for the local tokens and
builds a statistic row, which includes the histograms $H_{q,b}$ of Section~\ref{sec:prefill_algorithm}, the
protected loads, and the token count, $P(B+1)+1$ integers in total (about 8~KiB for $P=8$ and
$B=256$). Second, DeepEP's \texttt{notify\_dispatch}, which already exchanges each rank's token
counts, carries this row to all ranks together with the counts, both within and across nodes.
Third, once dispatch completes, a price kernel computes the prices of
Equation~\ref{eq:prefill_price} from the summed rows (Appendix~\ref{app:prefill_pipeline}).
Finally, the next MoE layer's selector applies Equation~\ref{eq:prefill_selection} with these
prices, taking gating weights from the original logits.

Both kernels are small and overlap with model execution (Figure~\ref{fig:rankcast_system}(a)). On
Qwen3-235B, the statistics and price kernels take 54 and 44\,$\mu$s per layer. The statistics
kernel uses one warp per token and overlaps with the current layer's expert selection and dispatch
setup. The price kernel uses a single 128-thread block, and its slack spans the current layer's
expert computation and the next layer's attention.

\paragraph{Expert-shedding pipeline for decode.}\label{sec:decode_implementation}
Decode also runs the four stages of Figure~\ref{fig:rankcast_system}(b) on a side stream. First,
each rank evaluates Equation~\ref{eq:cross_layer_prediction} for its local tokens and writes one
candidate row per token, with the token's Top-$L$ experts and their quality weights. Second, the
rows ride the token-count messages of DeepEP's low-latency dispatch to every rank, both within and
across nodes (Appendix~\ref{app:decode_payload}). Third, each rank runs the solver of
Section~\ref{sec:decode_algorithm} in a single CUDA block, which finishes during the current
layer's expert computation (Appendix~\ref{app:decode_cuda_layout}). Finally, the target layer
applies the mask by adding a large negative bias to the router score of every closed expert. Its
Top-$k$ therefore runs unchanged and never selects a closed expert. The whole path is CUDA-graph
compatible (Appendix~\ref{app:decode_abi_validation}).

\section{Evaluation}\label{sec:evaluation}

\subsection{Setup}\label{sec:eval_setup}

\paragraph{Models and deployment.} We evaluate RankCast on
Qwen3-235B-A22B-Instruct-2507~\citep{qwen3,qwen2025qwen3235b2507} in BF16 (94 MoE layers, 128
experts, $k=8$) and Qwen3.5-397B-A17B~\citep{qwen2026qwen35} (60 layers, 512 experts, $k=10$). Both run on two nodes of
$8\times$H800 connected over RoCE, served by SGLang 0.5.12~\citep{zheng2024sglang} with DeepEP 1.2.1~\citep{deepep2025}
and DeepGEMM~\citep{deepseek2025deepgemm}. Qwen3-235B runs prefill and decode on separate
nodes, each with TP$=8$ and EP$=8$. Qwen3.5-397B runs both phases in one instance across all 16
GPUs, with DP$=16$ attention and EP$=16$. Unless stated otherwise, RankCast uses $L=16$ candidates
($L=20$ on Qwen3.5-397B) and protects each token's top $p=2$ experts in both phases. Prefill uses
$B=256$ histogram bins on Qwen3-235B and $B=64$ on Qwen3.5-397B, and decode uses quality weights
without a depth cap.
Appendix~\ref{app:eval_extra} reports ablations of $p$, $D$, and $L$.

\paragraph{Workloads and baselines.} Prefill experiments measure input throughput on prompts from
ShareGPT~\citep{sharegpt2023} and CodeContests~\citep{li2022alphacode}, tiled or truncated to 4,096
tokens, and on 32,768-token windows of LongBench v2~\citep{bai2025longbenchv2}. Decode
experiments measure the median time per output token (TPOT) on WildChat~\citep{zhao2024wildchat},
ShareGPT, and CodeContests prompts at their natural lengths, and on ShareGPT prompts tiled to 1,024
tokens. Quality experiments run eight benchmarks on both models. We compare RankCast with online
EPLB~\citep{deepseekai2025eplb} in prefill, and with XShare~\citep{vankov2026xshare} and
OEA~\citep{oncescu2025opportunistic} in an offline decode comparison. Throughput and quality
changes are relative to native Top-$k$ routing on the same deployment, labeled Baseline in the
tables. Appendix~\ref{app:eval_protocol} lists all settings.

\subsection{Throughput}
\label{sec:eval_throughput}

\begin{table}[t]
    \centering
    \caption{Prefill results at three chunked-prefill sizes per model. Throughput gain is relative to SGLang's native Top-$k$ routing (Baseline). Imbalance is the max-to-mean token count across EP ranks. Bold marks the best value of each metric in each row. Online EPLB rebalances every 50 forward passes on ShareGPT and CodeContests; EPLB-R adds 8 redundant experts on Qwen3-235B and 16 on Qwen3.5-397B.}
    \label{tab:prefill}
    \small
    \setlength{\tabcolsep}{4pt}
    \begin{tabular}{llccccccc}
        \toprule
                     & Tokens   & \multicolumn{3}{c}{Throughput gain} & \multicolumn{4}{c}{Imbalance (max/mean)}                                                                                                  \\
        \cmidrule(lr){3-5}\cmidrule(lr){6-9}
        Workload     & per rank & EPLB                                & EPLB-R                                  & RankCast                    & Baseline & EPLB                    & EPLB-R                  & RankCast \\
        \midrule
        \multicolumn{9}{l}{\textit{Qwen3-235B}}                                                                                                                                                                                  \\
        ShareGPT     & 2K       & $+13.3\%$                           & $+7.0\%$                     & \boldmath$+16.4\%$        & 1.76     & 1.15                    & 1.41                    & \textbf{1.09}     \\
                     & 4K       & $+14.7\%$                           & $+8.0\%$                     & \boldmath$+18.5\%$                   & 1.75     & 1.13                    & 1.42                    & \textbf{1.08}     \\
                     & 8K       & $+13.8\%$                           & $+6.6\%$                     & \boldmath$+19.4\%$                   & 1.75     & 1.11                    & 1.46                    & \textbf{1.08}     \\
        \addlinespace[2pt]
        CodeContests & 2K       & $+12.7\%$                           & $+6.4\%$                     & \boldmath$+16.4\%$                   & 1.66     & 1.10                    & 1.35                    & \textbf{1.04}     \\
                     & 4K       & $+14.8\%$                           & $+7.0\%$                     & \boldmath$+18.8\%$                   & 1.66     & 1.09                    & 1.36                    & \textbf{1.04}     \\
                     & 8K       & $+14.2\%$                           & $+7.5\%$                     & \boldmath$+19.3\%$                   & 1.66     & 1.06                    & 1.39                    & \textbf{1.04}     \\
        \addlinespace[2pt]
        LongBench    & 2K       & \phantom{0}$-0.4\%$                 & $-1.2\%$                     & \boldmath\phantom{$0$}$+7.3\%$         & 1.66     & 1.60                    & 1.61                    & \textbf{1.16}     \\
                     & 4K       & \phantom{0}$+0.1\%$                 & $-0.2\%$                     & \boldmath\phantom{$0$}$+7.1\%$         & 1.65     & 1.58                    & 1.61                    & \textbf{1.15}     \\
                     & 8K       & \phantom{0}$+0.9\%$                 & $+0.3\%$                     & \boldmath\phantom{$0$}$+4.3\%$         & 1.47     & 1.39                    & 1.44                    & \textbf{1.10}     \\
        \midrule
        \multicolumn{9}{l}{\textit{Qwen3.5-397B}}                                                                                                                                                                                \\
        ShareGPT     & 4K       & \phantom{0}$+4.6\%$                 & $+4.4\%$                     & \boldmath$+10.8\%$                   & 1.51     & 1.14                    & 1.20                    & \textbf{1.09}     \\
                     & 8K       & \phantom{0}$+7.4\%$                 & $+7.1\%$                     & \boldmath$+11.2\%$                   & 1.50     & 1.11                    & 1.17                    & \textbf{1.07}     \\
                     & 16K      & \phantom{0}$+7.7\%$                 & $+7.3\%$   & \boldmath$+10.4\%$                   & 1.50     & \textbf{1.09}           & 1.13  & 1.11              \\
        \addlinespace[2pt]
        CodeContests & 4K       & \phantom{0}$+7.1\%$                 & $+6.5\%$                     & \boldmath$+13.9\%$                   & 1.55     & 1.08                    & 1.19                    & \textbf{1.04}     \\
                     & 8K       & \phantom{0}$+9.9\%$                 & $+9.1\%$                     & \boldmath$+14.6\%$                   & 1.55     & 1.06                    & 1.17                    & \textbf{1.03}     \\
                     & 16K      & $+10.4\%$         & $+9.7\%$   & \boldmath$+14.0\%$                   & 1.54     & 1.05  & 1.15  & \textbf{1.03}     \\
        \addlinespace[2pt]
        LongBench    & 8K       & \phantom{0}$-0.2\%$                 & $+0.2\%$                     & \boldmath\phantom{$0$}$+4.2\%$         & 1.26     & 1.15                    & 1.16                    & \textbf{1.05}     \\
                     & 16K      & \phantom{0}$+0.7\%$                 & ---                                     & \boldmath\phantom{$0$}$+4.4\%$         & 1.25     & 1.14                    & ---                     & \textbf{1.05}     \\
        \bottomrule
    \end{tabular}
\end{table}

\paragraph{Prefill.} Table~\ref{tab:prefill} reports the three workloads on both models. RankCast
lowers the imbalance across EP ranks to 1.03--1.16 and removes 76--95\% of the busiest rank's
excess load over the mean. On ShareGPT and CodeContests, input throughput rises by 16.4--19.4\% on
Qwen3-235B and 10.4--14.6\% on Qwen3.5-397B. On LongBench v2, it rises by 4.2--7.3\%, because the
32,768-token inputs make attention a larger share of each step. On ShareGPT and CodeContests,
online EPLB~\citep{deepseekai2025eplb} removes 73--91\% of the excess and raises input throughput
by 12.7--14.8\% on Qwen3-235B and by 4.6--10.4\% on Qwen3.5-397B. In these settings, RankCast
gains 2.7--6.8 points more than EPLB. It also leaves a lower imbalance, except on ShareGPT at 16K
tokens per rank on Qwen3.5-397B. Redundant experts lower the gain of EPLB on both models, because
the extra expert copies use GPU memory and therefore reduce the number of concurrent requests. On
LongBench v2, EPLB and EPLB-R remove only 6--43\% of the excess and change throughput by $-1.2$ to
$+0.9\%$.

To reach larger $P$ we replay the prices offline on captured scores. On Qwen3-235B prefill chunks,
native imbalance grows from 1.60 at $P=8$ to 5.30 at $P=64$ as each rank holds fewer experts, and
RankCast removes 53--70\% of the excess. On Qwen3.5-397B it removes 72\% of the excess at $P=16$
and 51\% at $P=128$, where native imbalance reaches 9.34 (Table~\ref{tab:ep_scaling} in
Appendix~\ref{app:eval_extra}).

\begin{table}[t]
    \centering
    \caption{The slow offset $\eta$ alone against full RankCast on Qwen3-235B prefill, one run per arm, with the clamp on $\eta$ relaxed to the full price range. Imbalance is the max-to-mean token count within one forward pass.}
    \label{tab:eta_ablation}
    \small
    \begin{tabular}{llcc}
        \toprule
        Workload          & Arm                       & Throughput gain & Imbalance \\
        \midrule
        ShareGPT, 4K      & Top-$k$                   & ---             & 1.75      \\
                          & $\eta$ alone              & $+14.7\%$       & 1.20      \\
                          & $\eta$ + per-batch prices & $+18.1\%$       & 1.08      \\
        \addlinespace[2pt]
        LongBench v2, 32K & Top-$k$                   & ---             & 1.66      \\
                          & $\eta$ alone              & $-0.8\%$        & 1.65      \\
                          & $\eta$ + per-batch prices & $+7.4\%$        & 1.16      \\
        \bottomrule
    \end{tabular}
\end{table}

\paragraph{Per-batch prices.} The slow per-rank offset $\eta$ of
Appendix~\ref{app:online_corrections} works like a training-time load bias. It accumulates across
forward passes and does not react within one. Table~\ref{tab:eta_ablation} runs it alone. On
ShareGPT at 4K, consecutive batches route similarly, so one standing offset fits them all and
recovers most of the gain, lowering imbalance from 1.75 to 1.20. On 32,768-token LongBench v2
inputs it moves imbalance only from 1.66 to 1.65 and costs 0.8\% of throughput. A standing offset
cannot follow imbalance that changes from batch to batch, while per-batch prices reach 1.16 and
$+7.4\%$.

\begin{table}[t]
    \centering
    \caption{Median decode TPOT (ms) with 1,024 output tokens per request. Qwen3-235B keeps up to 96 requests in flight, and Qwen3.5-397B up to 128. Prompt gives the median prompt length in tokens. $\Delta$ is the relative gain in $1/\mathrm{TPOT}$.}
    \label{tab:decode_throughput}
    \small
    \setlength{\tabcolsep}{4pt}
    \begin{tabular}{lrcccccc}
        \toprule
                        &        & \multicolumn{3}{c}{Qwen3-235B} & \multicolumn{3}{c}{Qwen3.5-397B} \\
        \cmidrule(lr){3-5}\cmidrule(lr){6-8}
        Workload        & Prompt & Baseline & RankCast & $\Delta$ & Baseline & RankCast & $\Delta$  \\
        \midrule
        WildChat        & 29     & 27.68    & 26.10    & $+6.1\%$ & 41.74    & 37.67    & $+10.8\%$ \\
        ShareGPT        & 91     & 27.69    & 26.11    & $+6.1\%$ & 40.76    & 37.48    & $+8.8\%$  \\
        ShareGPT, tiled & 1,024  & 27.59    & 25.58    & $+7.9\%$ & 37.53    & 34.16    & $+9.9\%$  \\
        CodeContests    & 3.5K   & 27.41    & 25.33    & $+8.2\%$ & 52.19    & 46.97    & $+11.1\%$ \\
        \bottomrule
    \end{tabular}
\end{table}

\paragraph{Decode.} Table~\ref{tab:decode_throughput} reports median decode TPOT on four workloads.
Decode shedding raises $1/\mathrm{TPOT}$ by 6.1--8.2\% on Qwen3-235B and by 8.8--11.1\% on
Qwen3.5-397B. Natural-length prompts still gain 6.1--11.1\%, so the gain does not depend on tiled
inputs. On Qwen3-235B, decode shedding lowers the mean number of active experts on the busiest rank
from 14.4 to 9.7 (Figure~\ref{fig:eval_decode_pred}(a) in Appendix~\ref{app:eval_protocol}). Across 11 configurations, median TPOT is
linear in this count, with a slope of 0.52\,ms per expert ($R^2=0.99$). The intercept of 20.1\,ms is
the part of TPOT that shedding cannot reduce.

\paragraph{Decode baselines.} XShare and OEA select experts from the true router scores of the whole
batch. Neither has an open-source implementation for expert parallelism. Running them on our
cross-layer prediction would therefore require changing their algorithms. We instead compare the
three methods offline on router scores captured from Qwen3-235B decode
(Table~\ref{tab:decode_baselines} in Appendix~\ref{app:eval_protocol}). XShare and OEA receive the
true scores, whereas RankCast uses its prediction and adds no collective. At the same $M$, XShare
and OEA still displace 1.9--6.0 times as much routing probability as RankCast.

\paragraph{Overhead.} Prefill balancing adds about 0.5\% to the prefill step time on both models.
Decode shedding adds 0.7\% to the decode step time on Qwen3-235B and 0.9\% on Qwen3.5-397B, mostly
through the prediction and masking kernels. Its solver takes 9--17\,$\mu$s per layer on Qwen3-235B
and 58\,$\mu$s on Qwen3.5-397B. The solver runs on a side stream during the layer's expert
computation, so almost none of this time is exposed.

\subsection{Quality}
\label{sec:eval_quality}

\begin{table}[t]
    \centering
    \caption{Quality change in percentage points, relative to native Top-$k$ routing. Prefill enables prefill balancing, Decode enables decode shedding, and Both enables the two together. Baseline gives the accuracy of native routing. Appendix~\ref{app:eval_protocol} lists the repeat counts.}
    \label{tab:quality}
    \small
    \setlength{\tabcolsep}{4pt}
    \begin{tabular}{lccccccc}
        \toprule
                         & \multicolumn{4}{c}{Qwen3-235B}     & \multicolumn{3}{c}{Qwen3.5-397B} \\
        \cmidrule(lr){2-5}\cmidrule(lr){6-8}
        Benchmark        & Baseline & Prefill & Decode  & Both    & Baseline & Prefill & Decode  \\
        \midrule
        MMLU-Redux 2.0   & 0.9105   & $-0.02$ & $-0.13$ & $-0.17$ & 0.9254   & $+0.18$ & $-0.01$ \\
        MATH-500         & 0.9813   & $-0.07$ & $-0.20$ & $+0.00$ & 0.9160   & $-0.07$ & $-0.30$ \\
        MMMLU            & 0.8198   & $-0.16$ & $-0.22$ & $-0.11$ & 0.8673   & $-0.04$ & $-0.25$ \\
        RULER            & 0.8787   & $-0.40$ & $-0.31$ & $-0.73$ & 0.9419   & $-0.20$ & $+0.02$ \\
        \midrule
        GPQA-Diamond     & 0.7621   & $-0.67$ & $-0.10$ & $-0.46$ & 0.8576   & $+0.03$ & $-0.39$ \\
        AIME 2025        & 0.6963   & $-0.32$ & $-0.02$ & $-0.30$ & 0.8707   & $+0.32$ & $-0.09$ \\
        IFEval strict    & 0.8848   & $-0.31$ & $-0.65$ & $-0.59$ & 0.8854   & $+0.02$ & $-0.19$ \\
        LiveCodeBench v5 & 0.7008   & $-0.43$ & $-0.59$ & $-0.84$ & 0.8467   & $-0.43$ & $-0.34$ \\
        \bottomrule
    \end{tabular}
\end{table}

Table~\ref{tab:quality} reports the accuracy change of prefill balancing, decode shedding, and
both together. On Qwen3-235B, prefill balancing lowers each of the four ten-repeat benchmarks by
0.3--0.7\,pp. Decode shedding changes no benchmark by more than 0.65\,pp. With both enabled, the
largest drop is 0.84\,pp, on LiveCodeBench. On Qwen3.5-397B no benchmark drops by more than
0.43\,pp. Appendix~\ref{app:eval_protocol} compares these changes with the spread between baseline
repeats.

\subsection{Prediction Accuracy}
\label{sec:eval_prediction}

\begin{table}[t]
    \centering
    \caption{Accuracy of the protected prefix and the protection it achieves at $p=2$, averaged over layers. The prefill column covers the 12 sampled MoE layers of Figure~\ref{fig:eval_decode_pred}(b) in Appendix~\ref{app:eval_protocol}; the decode columns cover every layer after the first, on the chat and code captures of Table~\ref{tab:decode_baselines}. Protection is measured after decode shedding. In prefill the selector uses the target layer's own scores, so the true Top-$p$ is exact.}
    \label{tab:prediction}
    \small
    \begin{tabular}{lccc}
        \toprule
                                                 & 235B prefill & 235B decode  & 397B decode  \\
        \midrule
        True top-1 $=$ predicted top-1           & 77.8\%       & 76--78\%     & 66\%         \\
        True top-1 $\in$ predicted top-2         & 91.4\%       & 92--93\%     & 84\%         \\
        True top-2 set $=$ predicted top-2 set   & 63.1\%       & 66\%         & 48\%         \\
        True top-2 $\subseteq$ predicted Top-$L$ & 99.7\%       & 99.8\%       & 98.0--98.6\% \\
        \midrule
        Tokens keeping their true top-1 expert   & exact        & 99.4--99.5\% & 96.7--98.5\% \\
        Tokens keeping both true top-2 experts   & exact        & 97.6--97.8\% & 87.9--94.5\% \\
        \bottomrule
    \end{tabular}
\end{table}

In Table~\ref{tab:prediction}, the predicted top-2 set matches the true one for 63.1\% of tokens in
235B prefill, 66\% in 235B decode, and 48\% in 397B decode. After shedding, 97.6--97.8\% of 235B
tokens and 87.9--94.5\% of 397B tokens keep both true top-2 experts. The true top-2 lies inside the
predicted Top-$L$ for 98.0--99.8\% of tokens, so a mis-ranked expert is still a candidate.
Mis-ranked pairs usually swap experts of near-equal weight. Prefill retention is exact by
construction. Qwen3.5-397B has a flatter router, so exact ranking is harder.

\section{Conclusion and Limitations}\label{sec:conclusion}

We presented RankCast, a unified framework for rank-level load balancing in expert-parallel MoE
inference. Prefill and decode share the same routing constraint, which retains each token's
predicted Top-$p$ experts. They differ only in the objective, which is the maximum per-rank token
load in prefill and the maximum active-expert count in decode. On Qwen3-235B-A22B and
Qwen3.5-397B-A17B, RankCast improves prefill input throughput by up to 19\% and reduces decode TPOT
by up to 10\%, at a quality cost below one percentage point.

RankCast changes which experts a token uses, so unlike expert placement or replication it is not
lossless. A deployment that needs the trained routing exactly should use placement instead.
RankCast acts on routing and placement acts on weights, so the two can be combined. We have not
measured them together.

We evaluate two models from one family. In both, the next layer's routing can be predicted from
that layer's own gate weights. Routers with group-limited or sigmoid gating remain untested. Our
experiments use at most 16 GPUs at EP$=16$. The kernels and the serving integration are not fully
optimized, so the overhead in Section~\ref{sec:eval_throughput} is an upper bound.

\section*{AI use statement}
We used generative AI tools to implement code: parts of the RankCast solver and its serving integration were AI-assisted. We additionally used them to edit the paper for readability and to write auxiliary scripts. We did not use them for the conceptual framework, hypotheses, experiment design, or interpretation of results; synthetic data, dataset cleaning, qualitative analysis, and translation are not applicable. The authors tested all AI-assisted code against the reference implementation of Appendix~\ref{app:implementation}, and take responsibility for the final content of this work.

\section*{Ethics statement}
This work improves the serving efficiency of existing, publicly released Mixture-of-Experts models
and introduces no new model, training procedure, or dataset. All experiments use public model
weights and benchmark datasets, involve no human subjects, and collect no personal data. Because
RankCast changes routing at inference time, we report the measured quality change of every
configuration, including degradations, so that deployers can weigh throughput against accuracy.

\section*{Reproducibility statement}
Section~\ref{sec:evaluation} and Appendix~\ref{app:eval_protocol} specify the model, serving stack, hardware, workloads, seeds, warmup and measurement windows, and repetition counts behind every reported number, and Appendix~\ref{app:eval_extra} reports the ablations of the protected prefix $p$, the depth cap $D$, the quality weights, and the candidate width $L$. The prefill solver is specified by Equations~\ref{eq:prefill_primal}--\ref{eq:prefill_selection}, with its dual derivation, capacity construction, and histogram error bound in Appendix~\ref{app:prefill_solver}; the decode solver by Equations~\ref{eq:decode_active_set_problem}--\ref{eq:decode_prefix_delta}, with its quality weights, closure order, repair step, and depth cap in Appendix~\ref{app:decode_solver}. Appendix~\ref{app:implementation} documents the kernels, the metadata exchange, the online corrections, the candidate-row layout, and the reference implementation against which the GPU solver was validated.

\bibliography{reference}

@inproceedings{zheng2024sglang,
  title     = {{SGLang}: Efficient Execution of Structured Language Model Programs},
  author    = {Zheng, Lianmin and Yin, Liangsheng and Xie, Zhiqiang and
               Sun, Chuyue and Huang, Jeff and Yu, Cody Hao and Cao, Shiyi and
               Kozyrakis, Christos and Stoica, Ion and Gonzalez, Joseph E. and
               Barrett, Clark and Sheng, Ying},
  booktitle = {Advances in Neural Information Processing Systems},
  volume    = {37},
  pages     = {62557--62583},
  year      = {2024},
  publisher = {Curran Associates, Inc.},
  doi       = {10.52202/079017-2000},
  url       = {https://proceedings.neurips.cc/paper_files/paper/2024/file/724be4472168f31ba1c9ac630f15dec8-Paper-Conference.pdf}
}

@misc{longcat,
  title        = {{LongCat-Flash} Technical Report},
  author       = {{Meituan LongCat Team}},
  year         = {2025},
  howpublished = {arXiv preprint arXiv:2509.01322},
  eprint       = {2509.01322},
  archivePrefix= {arXiv},
  primaryClass = {cs.CL},
  url          = {https://arxiv.org/abs/2509.01322}
}

@misc{deepep2025,
  title        = {{DeepEP}: an efficient expert-parallel communication library},
  author       = {Zhao, Chenggang and Zhou, Shangyan and Zhang, Liyue and Deng, Chengqi and
                  Xu, Zhean and Liu, Yuxuan and Yu, Kuai and Li, Jiashi and Zhao, Liang},
  year         = {2025},
  howpublished = {Software repository},
  url          = {https://github.com/deepseek-ai/DeepEP}
}

@misc{jiang2024mixtral,
  title        = {Mixtral of Experts},
  author       = {Jiang, Albert Q. and Sablayrolles, Alexandre and Roux, Antoine and Mensch, Arthur and Savary, Blanche and Bamford, Chris and Chaplot, Devendra Singh and de las Casas, Diego and Hanna, Emma Bou and Bressand, Florian and others},
  year         = {2024},
  howpublished = {arXiv preprint arXiv:2401.04088},
  eprint       = {2401.04088},
  archiveprefix = {arXiv},
  primaryclass = {cs.LG},
  url          = {https://arxiv.org/abs/2401.04088}
}

@misc{qwen3,
    title        = {{Qwen3} Technical Report},
    author       = {An Yang and Anfeng Li and Baosong Yang and Beichen Zhang and Binyuan Hui and Bo Zheng and Bowen Yu and Chang Gao and Chengen Huang and Chenxu Lv and Chujie Zheng and Dayiheng Liu and Fan Zhou and Fei Huang and Feng Hu and Hao Ge and Haoran Wei and Huan Lin and Jialong Tang and Jian Yang and Jianhong Tu and Jianwei Zhang and Jianxin Yang and Jiaxi Yang and Jing Zhou and Jingren Zhou and Junyang Lin and Kai Dang and Keqin Bao and Kexin Yang and Le Yu and Lianghao Deng and Mei Li and Mingfeng Xue and Mingze Li and Pei Zhang and Peng Wang and Qin Zhu and Rui Men and Ruize Gao and Shixuan Liu and Shuang Luo and Tianhao Li and Tianyi Tang and Wenbiao Yin and Xingzhang Ren and Xinyu Wang and Xinyu Zhang and Xuancheng Ren and Yang Fan and Yang Su and Yichang Zhang and Yinger Zhang and Yu Wan and Yuqiong Liu and Zekun Wang and Zeyu Cui and Zhenru Zhang and Zhipeng Zhou and Zihan Qiu},
    year         = {2025},
    howpublished = {arXiv preprint arXiv:2505.09388},
    eprint       = {2505.09388},
    archiveprefix = {arXiv},
    primaryclass = {cs.CL},
    url          = {https://arxiv.org/abs/2505.09388}
}

@misc{qwen2026qwen35,
  title        = {{Qwen3.5}: Towards Native Multimodal Agents},
  author       = {{Qwen Team}},
  year         = {2026},
  month        = feb,
  howpublished = {Blog post},
  url          = {https://qwen.ai/blog?id=qwen3.5},
  note         = {Model card: \url{https://huggingface.co/Qwen/Qwen3.5-397B-A17B}}
}

@article{fedus2022switch,
  title   = {{Switch Transformers}: Scaling to Trillion Parameter Models with Simple and Efficient Sparsity},
  author  = {Fedus, William and Zoph, Barret and Shazeer, Noam},
  journal = {Journal of Machine Learning Research},
  volume  = {23},
  number  = {120},
  pages   = {1--39},
  year    = {2022},
  url     = {https://jmlr.org/papers/v23/21-0998.html}
}

@inproceedings{lewis2021base,
  title     = {{BASE} Layers: Simplifying Training of Large, Sparse Models},
  author    = {Lewis, Mike and Bhosale, Shruti and Dettmers, Tim and Goyal, Naman and Zettlemoyer, Luke},
  booktitle = {Proceedings of the 38th International Conference on Machine Learning},
  series    = {Proceedings of Machine Learning Research},
  volume    = {139},
  pages     = {6265--6274},
  year      = {2021},
  url       = {https://proceedings.mlr.press/v139/lewis21a.html}
}

@inproceedings{zhou2022expertchoice,
  title     = {Mixture-of-Experts with Expert Choice Routing},
  author    = {Zhou, Yanqi and Lei, Tao and Liu, Hanxiao and Du, Nan and Huang, Yanping and Zhao, Vincent and Dai, Andrew M. and Chen, Zhifeng and Le, Quoc V. and Laudon, James},
  booktitle = {Advances in Neural Information Processing Systems},
  volume    = {35},
  pages     = {7103--7114},
  year      = {2022},
  url       = {https://proceedings.neurips.cc/paper_files/paper/2022/hash/2f00ecd787b432c1d36f3de9800728eb-Abstract-Conference.html}
}

@misc{wang2024auxlossfree,
  title         = {Auxiliary-Loss-Free Load Balancing Strategy for Mixture-of-Experts},
  author        = {Wang, Lean and Gao, Huazuo and Zhao, Chenggang and Sun, Xu and Dai, Damai},
  year          = {2024},
  howpublished  = {arXiv preprint arXiv:2408.15664},
  eprint        = {2408.15664},
  archiveprefix = {arXiv},
  primaryclass  = {cs.LG},
  url           = {https://arxiv.org/abs/2408.15664}
}

@misc{deepseekai2024deepseekv3,
  title         = {{DeepSeek-V3} Technical Report},
  author        = {{DeepSeek-AI}},
  year          = {2024},
  howpublished  = {arXiv preprint arXiv:2412.19437},
  eprint        = {2412.19437},
  archiveprefix = {arXiv},
  primaryclass  = {cs.CL},
  url           = {https://arxiv.org/abs/2412.19437}
}

@misc{deepseekai2026deepseekv4,
  title         = {{DeepSeek-V4}: Towards Highly Efficient Million-Token Context Intelligence},
  author        = {{DeepSeek-AI}},
  year          = {2026},
  howpublished  = {arXiv preprint arXiv:2606.19348},
  eprint        = {2606.19348},
  archiveprefix = {arXiv},
  primaryclass  = {cs.CL},
  url           = {https://arxiv.org/abs/2606.19348}
}

@misc{kimiteam2026k3,
  title         = {{Kimi K3}: Open Frontier Intelligence},
  author        = {{Kimi Team}},
  year          = {2026},
  howpublished  = {arXiv preprint arXiv:2607.24653},
  eprint        = {2607.24653},
  archiveprefix = {arXiv},
  primaryclass  = {cs.CL},
  url           = {https://arxiv.org/abs/2607.24653}
}

@inproceedings{li2023lina,
  title     = {Accelerating Distributed {MoE} Training and Inference with {Lina}},
  author    = {Li, Jiamin and Jiang, Yimin and Zhu, Yibo and Wang, Cong and Xu, Hong},
  booktitle = {2023 USENIX Annual Technical Conference (USENIX ATC 23)},
  pages     = {945--959},
  year      = {2023},
  url       = {https://www.usenix.org/conference/atc23/presentation/li-jiamin}
}

@misc{deepseekai2025eplb,
  title        = {{EPLB}: Expert Parallelism Load Balancer},
  author       = {{DeepSeek-AI}},
  year         = {2025},
  howpublished = {Software repository},
  url          = {https://github.com/deepseek-ai/EPLB},
  note         = {Accessed 2026-08-09}
}

@misc{go2025moetuner,
  title         = {{MoETuner}: Optimized Mixture of Expert Serving with Balanced Expert Placement and Token Routing},
  author        = {Go, Seokjin and Mahajan, Divya},
  year          = {2025},
  howpublished  = {arXiv preprint arXiv:2502.06643},
  eprint        = {2502.06643},
  archiveprefix = {arXiv},
  primaryclass  = {cs.LG},
  url           = {https://arxiv.org/abs/2502.06643}
}

@inproceedings{li2026semantic,
  title     = {Semantic Parallelism: Redefining Efficient {MoE} Inference via Model-Data Co-Scheduling},
  author    = {Li, Yan and Zhang, Zhenyu and Wang, Zhengang and Chen, Pengfei and Zheng, Pengfei},
  booktitle = {The Fourteenth International Conference on Learning Representations},
  year      = {2026},
  url       = {https://openreview.net/forum?id=MSHPrMpIHZ}
}

@misc{han2025gracemoe,
  title         = {{GRACE-MoE}: Grouping and Replication with Locality-Aware Routing for Efficient Distributed {MoE} Inference},
  author        = {Han, Yu and Pan, Lehan and Peng, Jie and Tao, Ziyang and Zhu, Hanqi and Zhang, Wuyang and Zhang, Yanyong},
  year          = {2025},
  howpublished  = {arXiv preprint arXiv:2509.25041},
  eprint        = {2509.25041},
  archiveprefix = {arXiv},
  primaryclass  = {cs.DC},
  url           = {https://arxiv.org/abs/2509.25041}
}

@inproceedings{zhao2026craft,
  title     = {{CRAFT}: Fine-Grained Cost-Aware Expert Replication for Efficient Mixture-of-Experts Serving},
  author    = {Zhao, Adrian and Cai, Zhenkun and Song, Zhenyu and Yu, Lingfan and Fan, Haozheng and Wu, Jun and Wang, Yida and Vijaykumar, Nandita},
  booktitle = {Proceedings of Machine Learning and Systems},
  volume    = {8},
  pages     = {689--700},
  year      = {2026},
  url       = {https://proceedings.mlsys.org/paper_files/paper/2026/hash/3a7f9e485845dac27423375c934cb4db-Abstract-Conference.html}
}

@inproceedings{yang2026libra,
  title     = {Libra: Effective yet Efficient Load Balancing for Large-Scale {MoE} Inference},
  author    = {Yang, Jaehoon and Kim, Yushin and Moon, Seokwon and Park, Yeonhong and Lee, Jae W.},
  booktitle = {The Fourteenth International Conference on Learning Representations},
  year      = {2026},
  url       = {https://openreview.net/forum?id=WhxNwgGkAS}
}

@misc{zhu2026probe,
  title         = {{PROBE}: Co-Balancing Computation and Communication in {MoE} Inference via Real-Time Predictive Prefetching},
  author        = {Zhu, Qianchao and Ye, Xucheng and Liu, Yuliang and Ouyang, Haodong and Song, Chengru},
  year          = {2026},
  howpublished  = {arXiv preprint arXiv:2602.00509},
  eprint        = {2602.00509},
  archiveprefix = {arXiv},
  primaryclass  = {cs.DC},
  url           = {https://arxiv.org/abs/2602.00509}
}

@misc{moonshotai2026moonep,
  title        = {{MoonEP}: A Perfectly Balanced Expert Parallelism Library via Dynamic Redundant Experts},
  author       = {Chen, Yutian and Li, Cong and Wang, Yucheng and Wei, Ming},
  year         = {2026},
  howpublished = {Software repository},
  url          = {https://github.com/MoonshotAI/MoonEP},
  note         = {Accessed 2026-08-09}
}

@misc{wei2026ultraep,
  title         = {{UltraEP}: Unleash {MoE} Training and Inference on Rack-Scale Nodes with Near-Optimal Load Balancing},
  author        = {Wei, Xinming and Jin, Chao and Dai, Tuo and Zhong, Yinmin and Yu, Shan and Yang, Chengxu and Wu, Bingyang and Zhang, Zili and Mai, Jing and Zhu, Qianchao and Li, Zhouyang and Liu, Yuliang and Luo, Guojie},
  year          = {2026},
  howpublished  = {arXiv preprint arXiv:2606.04101},
  eprint        = {2606.04101},
  archiveprefix = {arXiv},
  primaryclass  = {cs.DC},
  url           = {https://arxiv.org/abs/2606.04101}
}

@inproceedings{he2026capacity,
  title     = {Capacity-Aware Inference: Mitigating the Straggler Effect in Mixture of Experts},
  author    = {He, Shwai and Cai, Weilin and Huang, Jiayi and Li, Ang},
  booktitle = {The Fourteenth International Conference on Learning Representations},
  year      = {2026},
  url       = {https://openreview.net/forum?id=LuYFpySWA2}
}

@misc{shahout2025laser,
  title         = {From Score Distributions to Balance: Plug-and-Play Mixture-of-Experts Routing},
  author        = {Shahout, Rana and Cai, Colin and Du, Yilun and Yu, Minlan and Mitzenmacher, Michael},
  year          = {2025},
  howpublished  = {arXiv preprint arXiv:2510.03293},
  eprint        = {2510.03293},
  archiveprefix = {arXiv},
  primaryclass  = {cs.LG},
  url           = {https://arxiv.org/abs/2510.03293}
}

@inproceedings{wu2026sere,
  title     = {{SERE}: Similarity-based Expert Re-routing for Efficient Batch Decoding in {MoE} Models},
  author    = {Wu, Juntong and Cheng, Jialiang and Lv, Fuyu and Ou, Dan and Yuan, Li},
  booktitle = {The Fourteenth International Conference on Learning Representations},
  year      = {2026},
  url       = {https://openreview.net/forum?id=98IxaUQtMY}
}

@inproceedings{oncescu2025opportunistic,
  title     = {Opportunistic Expert Activation: Batch-Aware Expert Routing for Faster Decode Without Retraining},
  author    = {Oncescu, Costin-Andrei and Wu, Qingyang and Chung, Wai Tong and Wu, Tsai-chuan and Gopal, Bryan and Wang, Junxiong and Dao, Tri and Athiwaratkun, Ben},
  booktitle = {The Forty-Third International Conference on Machine Learning},
  year      = {2026},
  url       = {https://openreview.net/forum?id=bSLIDztFwx}
}

@misc{gupta2024lynx,
  title         = {{Lynx}: Enabling Efficient {MoE} Inference through Dynamic Batch-Aware Expert Selection},
  author        = {Gupta, Vima and Ju, Jae Hyung and Sinha, Kartik and Gavrilovska, Ada and Iyer, Anand Padmanabha},
  year          = {2024},
  howpublished  = {arXiv preprint arXiv:2411.08982},
  eprint        = {2411.08982},
  archiveprefix = {arXiv},
  primaryclass  = {cs.LG},
  url           = {https://arxiv.org/abs/2411.08982}
}

@misc{vankov2026xshare,
  title         = {{XShare}: Collaborative in-Batch Expert Sharing for Faster {MoE} Inference},
  author        = {Vankov, Daniil and Ivkin, Nikita and Ulrich, Kyle and Song, Xiang and Khetan, Ashish and Karypis, George},
  year          = {2026},
  howpublished  = {arXiv preprint arXiv:2602.07265},
  eprint        = {2602.07265},
  archiveprefix = {arXiv},
  primaryclass  = {cs.LG},
  url           = {https://arxiv.org/abs/2602.07265}
}

@misc{sharegpt2023,
  title        = {{ShareGPT}},
  author       = {{ShareGPT}},
  year         = {2023},
  howpublished = {Conversation-sharing platform},
  url          = {https://sharegpt.com/}
}

@inproceedings{zhao2024wildchat,
  title     = {{WildChat}: {1M} {ChatGPT} Interaction Logs in the Wild},
  author    = {Zhao, Wenting and Ren, Xiang and Hessel, Jack and Cardie, Claire and Choi, Yejin and Deng, Yuntian},
  booktitle = {The Twelfth International Conference on Learning Representations},
  year      = {2024}
}

@inproceedings{qin2025mooncake,
  title     = {Mooncake: Trading More Storage for Less Computation --- A {KVCache}-centric Architecture for Serving {LLM} Chatbot},
  author    = {Qin, Ruoyu and Li, Zheming and He, Weiran and Cui, Jialei and Ren, Feng and Zhang, Mingxing and Wu, Yongwei and Zheng, Weimin and Xu, Xinran},
  booktitle = {23rd USENIX Conference on File and Storage Technologies (FAST 25)},
  pages     = {155--170},
  year      = {2025},
  publisher = {USENIX Association},
  url       = {https://www.usenix.org/conference/fast25/presentation/qin}
}

@inproceedings{bai2025longbenchv2,
  title     = {{LongBench v2}: Towards Deeper Understanding and Reasoning on Realistic Long-context Multitasks},
  author    = {Bai, Yushi and Tu, Shangqing and Zhang, Jiajie and Peng, Hao and Wang, Xiaozhi and Lv, Xin and Cao, Shulin and Xu, Jiazheng and Hou, Lei and Dong, Yuxiao and Tang, Jie and Li, Juanzi},
  booktitle = {Proceedings of the 63rd Annual Meeting of the Association for Computational Linguistics (Volume 1: Long Papers)},
  pages     = {3639--3664},
  year      = {2025},
  doi       = {10.18653/v1/2025.acl-long.183},
  url       = {https://aclanthology.org/2025.acl-long.183/}
}

@inproceedings{gema2025mmlu,
  title     = {Are We Done with {MMLU}?},
  author    = {Gema, Aryo Pradipta and Leang, Joshua Ong Jun and Hong, Giwon and Devoto, Alessio and Mancino, Alberto Carlo Maria and Saxena, Rohit and He, Xuanli and Zhao, Yu and Du, Xiaotang and Ghasemi Madani, Mohammad Reza and Barale, Claire and McHardy, Robert and Harris, Joshua and Kaddour, Jean and Van Krieken, Emile and Minervini, Pasquale},
  booktitle = {Proceedings of the 2025 Conference of the Nations of the Americas Chapter of the Association for Computational Linguistics: Human Language Technologies (Volume 1: Long Papers)},
  pages     = {5069--5096},
  year      = {2025},
  doi       = {10.18653/v1/2025.naacl-long.262},
  url       = {https://aclanthology.org/2025.naacl-long.262/}
}

@inproceedings{hendrycks2021math,
  title     = {Measuring Mathematical Problem Solving With the {MATH} Dataset},
  author    = {Hendrycks, Dan and Burns, Collin and Kadavath, Saurav and Arora, Akul and Basart, Steven and Tang, Eric and Song, Dawn and Steinhardt, Jacob},
  booktitle = {Proceedings of the Neural Information Processing Systems Track on Datasets and Benchmarks},
  volume    = {1},
  year      = {2021},
  url       = {https://datasets-benchmarks-proceedings.neurips.cc/paper/2021/hash/be83ab3ecd0db773eb2dc1b0a17836a1-Abstract-round2.html}
}

@inproceedings{lightman2023verify,
  title     = {Let's Verify Step by Step},
  author    = {Lightman, Hunter and Kosaraju, Vineet and Burda, Yura and Edwards, Harri and Baker, Bowen and Lee, Teddy and Leike, Jan and Schulman, John and Sutskever, Ilya and Cobbe, Karl},
  booktitle = {The Twelfth International Conference on Learning Representations},
  year      = {2024},
  url       = {https://openreview.net/forum?id=v8L0pN6EOi}
}

@misc{maa2025aime,
  title        = {2025 {American Invitational Mathematics Examination} ({AIME}~{I} and~{II})},
  author       = {{Mathematical Association of America}},
  year         = {2025},
  howpublished = {Annual mathematics competition},
  url          = {https://maa.org/maa-invitational-competitions/}
}

@misc{openai2024mmmlu,
  title        = {{Multilingual Massive Multitask Language Understanding} ({MMMLU})},
  author       = {{OpenAI}},
  year         = {2024},
  howpublished = {Dataset},
  url          = {https://huggingface.co/datasets/openai/MMMLU}
}

@inproceedings{hsieh2024ruler,
  title     = {{RULER}: What's the Real Context Size of Your Long-Context Language Models?},
  author    = {Hsieh, Cheng-Ping and Sun, Simeng and Kriman, Samuel and Acharya, Shantanu and Rekesh, Dima and Jia, Fei and Zhang, Yang and Ginsburg, Boris},
  booktitle = {First Conference on Language Modeling},
  year      = {2024},
  url       = {https://openreview.net/forum?id=kIoBbc76Sy}
}

@inproceedings{rein2024gpqa,
  title     = {{GPQA}: A Graduate-Level {Google}-Proof {Q\&A} Benchmark},
  author    = {Rein, David and Hou, Betty Li and Stickland, Asa Cooper and Petty, Jackson and Pang, Richard Yuanzhe and Dirani, Julien and Michael, Julian and Bowman, Samuel R.},
  booktitle = {First Conference on Language Modeling},
  year      = {2024},
  url       = {https://openreview.net/forum?id=Ti67584b98}
}

@misc{zhou2023ifeval,
  title         = {Instruction-Following Evaluation for Large Language Models},
  author        = {Zhou, Jeffrey and Lu, Tianjian and Mishra, Swaroop and Brahma, Siddhartha and Basu, Sujoy and Luan, Yi and Zhou, Denny and Hou, Le},
  year          = {2023},
  howpublished  = {arXiv preprint arXiv:2311.07911},
  eprint        = {2311.07911},
  archiveprefix = {arXiv},
  primaryclass  = {cs.CL},
  url           = {https://arxiv.org/abs/2311.07911}
}

@inproceedings{jain2024livecodebench,
  title     = {{LiveCodeBench}: Holistic and Contamination Free Evaluation of Large Language Models for Code},
  author    = {Jain, Naman and Han, King and Gu, Alex and Li, Wen-Ding and Yan, Fanjia and Zhang, Tianjun and Wang, Sida and Solar-Lezama, Armando and Sen, Koushik and Stoica, Ion},
  booktitle = {The Thirteenth International Conference on Learning Representations},
  year      = {2025},
  url       = {https://openreview.net/forum?id=chfJJYC3iL}
}

@misc{qwen2025qwen3235b2507,
  title        = {{Qwen3-235B-A22B-Instruct-2507}},
  author       = {{Qwen Team}},
  year         = {2025},
  howpublished = {Model card},
  url          = {https://huggingface.co/Qwen/Qwen3-235B-A22B-Instruct-2507}
}

@misc{deepseek2025deepgemm,
  title        = {{DeepGEMM}: clean and efficient {BLAS} kernel library on {GPU}},
  author       = {Zhao, Chenggang and Xu, Zhean and Zhao, Liang and Li, Jiashi and Xu, Chenhao and Xu, Anyi and Liu, Shengyu and Zhou, Kexing and Yu, Kuai},
  year         = {2025},
  howpublished = {Software repository},
  url          = {https://github.com/deepseek-ai/DeepGEMM}
}

@misc{livecodebench2025releasev5,
  title        = {{LiveCodeBench release v5}},
  author       = {{LiveCodeBench Team}},
  year         = {2025},
  howpublished = {Official benchmark release},
  url          = {https://github.com/LiveCodeBench/LiveCodeBench}
}

@article{li2022alphacode,
  title   = {Competition-Level Code Generation with {AlphaCode}},
  author  = {Li, Yujia and Choi, David and Chung, Junyoung and Kushman, Nate and Schrittwieser, Julian and Leblond, R{\'e}mi and Eccles, Tom and Keeling, James and Gimeno, Felix and Dal Lago, Agustin and others},
  journal = {Science},
  volume  = {378},
  number  = {6624},
  pages   = {1092--1097},
  year    = {2022},
  doi     = {10.1126/science.abq1158}
}
\bibliographystyle{iclr2027_conference}

\appendix

\section{Prefill Solver Details}
\label{app:prefill_solver}

\subsection{Dual Derivation of the Rank Prices}
\label{app:price_derivation}

\paragraph{Exact LP relaxation.}
Let slot $(i,q,j)$ be \emph{valid} if $a_{i,q,j}>-\infty$, and let $U_q$ denote the number of valid
slots on rank $q$. Dropping the padded slots and relaxing $x_{i,q,j}$ to $[0,1]$ gives
\begin{equation}
\label{eq:app_rankcast_relaxation}
\max_{\mathbf{x}}\ \sum_{i,q,j}a_{i,q,j}x_{i,q,j}
\quad\mathrm{s.t.}\quad
\sum_{q,j}x_{i,q,j}=d\ \ \forall i,
\qquad
\sum_{i,j}x_{i,q,j}=c_q\ \ \forall q,
\qquad
0\leq x_{i,q,j}\leq1.
\end{equation}
This is a transportation problem. Token $i$ supplies $d$ units, rank $q$ demands $c_q$ units, and
each valid slot is an edge of capacity one. Its constraint matrix, including the bounds
$x_{i,q,j}\leq1$, is totally unimodular, and $d$ and $c_q$ are integers. Every vertex of the
feasible polytope is therefore integral. Whenever a feasible assignment exists, the relaxation has
an integral optimum with the same value as Equation~\ref{eq:prefill_primal}.

\paragraph{Dual and block updates.}
Let the threshold $\alpha_i$ and the price $\beta_q$ be the multipliers of the token and rank
constraints, and let $[z]_{+}=\max(z,0)$. The dual function is
\begin{equation}
\label{eq:app_rankcast_dual}
g(\boldsymbol{\alpha},\boldsymbol{\beta})
=
d\sum_i\alpha_i+\sum_qc_q\beta_q+\sum_{i,q,j}\bigl[a_{i,q,j}-\alpha_i-\beta_q\bigr]_{+}.
\end{equation}
By complementary slackness, an optimal assignment selects every slot whose adjusted score
$a_{i,q,j}-\beta_q$ exceeds $\alpha_i$ and no slot whose adjusted score falls below it. Each token
therefore takes its top-$d$ adjusted scores, as in Equation~\ref{eq:prefill_selection}.

For fixed $\boldsymbol{\beta}$, let $v_{i,(m)}^\downarrow$ denote the $m$-th largest adjusted score
of token $i$. Any threshold in
\begin{equation}
\label{eq:app_alpha_interval}
\alpha_i\in\bigl[v_{i,(d+1)}^\downarrow,\,v_{i,(d)}^\downarrow\bigr]
\end{equation}
minimizes $g$ over $\alpha_i$. For fixed $\boldsymbol{\alpha}$, let $w_{q,(m)}^\downarrow$ denote the
$m$-th largest residual score on rank $q$, across all source ranks. Any price in
\begin{equation}
\label{eq:app_beta_interval}
\beta_q\in\bigl[w_{q,(c_q+1)}^\downarrow,\,w_{q,(c_q)}^\downarrow\bigr]
\end{equation}
minimizes $g$ over $\beta_q$. Starting from $\boldsymbol{\beta}=\mathbf{0}$ and taking the midpoints
of both intervals gives Equation~\ref{eq:prefill_coordinate_updates}. If $c_q=0$ or $c_q=U_q$, the price interval becomes one-sided:
\begin{equation}
\label{eq:app_capacity_boundaries}
\beta_q\geq w_{q,(1)}^\downarrow\ \ \text{if } c_q=0,
\qquad
\beta_q\leq w_{q,(U_q)}^\downarrow\ \ \text{if } c_q=U_q.
\end{equation}

\paragraph{One sweep and centering.}
One sweep does not in general reach the dual optimum, so rank $q$ need not receive exactly $c_q$
flexible assignments. Each further sweep would need another histogram exchange, so RankCast runs
one sweep. Because $\sum_qc_q=Nd$, raising every $\beta_q$ and lowering every $\alpha_i$ by the same
amount leaves $g$ unchanged. Equation~\ref{eq:prefill_price} fixes this shift by subtracting the
mean price.

\subsection{Flexible Target Loads}
\label{app:capacity}

Let $C_q$ denote the balanced total load of rank $q$. The $Nk$ assignments are split as evenly as
possible across ranks:
\begin{equation}
\label{eq:app_total_capacity}
C_q=\left\lfloor Nk/P\right\rfloor+\mathbf{1}\!\left[q\leq Nk\bmod P\right],
\qquad
\sum_{q=1}^{P}C_q=Nk.
\end{equation}
Every rank obtains $N$ from the token counts of the statistic rows, so all ranks compute the same
$C_q$. Let $f_q$ denote the protected load of rank $q$, summed over all source ranks. Every token
has $p$ protected experts, so the targets $C_q-f_q$ sum to $Nd$. A target can be negative or exceed
$U_q$, and clipping it to $[0,U_q]$ alone can change the sum. RankCast therefore subtracts a shared
offset $\mu$ before clipping:
\begin{equation}
\label{eq:app_capacity_waterfill}
z_q^{\star}(\mu)
=
\min\!\left\{U_q,\max\!\left(0,C_q-f_q-\mu\right)\right\},
\qquad
\sum_q z_q^{\star}(\mu)=Nd.
\end{equation}
Every token has at least $d$ valid slots, so $\sum_qU_q\geq Nd$. The clipped sum falls continuously
from $\sum_qU_q$ to $0$ as $\mu$ grows, so bisection finds $\mu$. The resulting $\mathbf z^\star$ is
the point closest to $\mathbf C-\mathbf f$ that satisfies the bounds and the sum. RankCast rounds
each $z_q^\star$ down and gives the remaining units, one each, to the unsaturated ranks with the
largest fractional parts. The integer targets therefore satisfy $0\leq c_q\leq U_q$ and
$\sum_qc_q=Nd$. These bounds do not guarantee a feasible assignment, because several ranks can
compete for the slots of the same tokens. The prices remain defined in that case.

\subsection{Accuracy of the Histogram Estimate}
\label{app:histogram_error}

The histograms use $B=256$ equal-width bins on $[-W,W]$ with $W=4$, so each bin has width
$\Delta\tau=2W/B$. Scores outside this range are counted in the first or the last bin. For rank
$q$, the price kernel accumulates bin counts from the top until the count reaches $c_q$. Let $b$
denote the bin where it stops, and let $H_{q,>b}$ denote the count above this bin. The estimate
interpolates linearly inside bin $b$:
\begin{equation}
\label{eq:app_quantile_interp}
Q_q(c_q)=\tau_{b+1}-\Delta\tau\,\frac{c_q-H_{q,>b}}{H_{q,b}}.
\end{equation}
Integer counts sum exactly, so bin $b$ contains the exact order statistic $w_{q,(c_q)}^\downarrow$
whenever $b$ is not an edge bin. Both values then lie in the same bin:
\begin{equation}
\label{eq:app_quantile_error}
\left|Q_q(c_q)-w_{q,(c_q)}^\downarrow\right|\leq\Delta\tau.
\end{equation}
The upper endpoint $w_{q,(c_q)}^\downarrow$ of Equation~\ref{eq:app_beta_interval} is an exact
block minimizer, like the midpoint of Equation~\ref{eq:prefill_coordinate_updates}. Each price is
therefore within $\Delta\tau$ of an exact update. Centering shifts every price equally and does not
change routing. A token keeps its exact selection whenever the gap between its $d$-th and
$(d+1)$-th price-adjusted scores exceeds $2\Delta\tau$. If $c_q=0$, the kernel sets the price just
above $W$. If $c_q=U_q$, it sets the price just below $-W$. These values realize
Equation~\ref{eq:app_capacity_boundaries} within the histogram range.

\section{Decode Solver Details}
\label{app:decode_solver}

\subsection{Quality Weights and Closure Order}
\label{app:decode_prefix_details}

\paragraph{Quality weights.}
Let $\hat s_{i,e}$ denote the predicted score of expert $e$ for token $i$, and let
$\hat s_{i,(m)}^\downarrow$ denote the $m$-th largest predicted score of token $i$. Let
$u_{i,e}=\exp(\hat s_{i,e}-\hat s_{i,(1)}^\downarrow)$, and define $u_{i,(m)}^\downarrow$ likewise.
Closing an expert $e$ of the predicted Top-$k$ approximately hands its routing weight to the
$(k+1)$-th expert. The quality weight is this loss, normalized by the loss for the top expert and
rounded up to a multiple of $1/3$:
\begin{equation}
\label{eq:decode_quality_weight}
\omega_{i,e}=\frac13\left\lceil\frac{3\,\bigl(u_{i,e}-u_{i,(k+1)}^\downarrow\bigr)}{1-u_{i,(k+1)}^\downarrow}\right\rceil
\quad\text{for } e\in\operatorname{Top}_k(\hat s_i),
\end{equation}
and $\omega_{i,e}=0$ for the other candidates. Each weight therefore takes one of four values and
fits in two bits. Without quality weights, the solver sets every $\omega_{i,e}$ to one.

\paragraph{Closure order.}
Ties in the closure cost are broken by the number of tokens that list the expert as a candidate,
fewest first, because closing such an expert reduces the slack of fewer tokens. Remaining ties are
broken by expert identifier, so every rank computes the same order.

\paragraph{Closure set.}
Let $\mathcal R(\Delta)=\{e\mid\operatorname{lev}(e)\leq\Delta\}$ denote the experts that
Section~\ref{sec:decode_algorithm} closes to lower $M$ by $\Delta$. For $\Delta\leq\Delta^\star$,
closing $\mathcal R(\Delta)$ lowers $M$ by exactly $\Delta$ and keeps every token's coverage.
$\Delta^\star$ is the largest reduction when every rank closes only its cheapest closable experts.
Closing other experts may allow a larger reduction, so the solver does not guarantee an optimum of
Equation~\ref{eq:decode_active_set_problem}. The solver works on predicted scores, so these
guarantees hold for the predicted sets $\mathcal F_i$ and $\mathcal C_i$.

\subsection{Repair Step}
\label{app:decode_frontier_repair}

$\Delta^\star=0$ in two cases. Either a busiest rank has no closable expert, or some token would
lose more than its slack if every busiest rank closed its cheapest closable expert. In the first
case, no closure can lower $M$. In the second case, closing other experts on the busiest ranks may
still lower $M$ by one. The repair step searches for one such expert per busiest rank:
\begin{enumerate}
    \item Each busiest rank $r$ starts with index $j_r=1$.
    \item Let $\mathcal U=\{e_{r,j_r}\mid A_r=M\}$ denote the proposal. A token $i$ \emph{conflicts}
          with $\mathcal U$ if $|\mathcal C_i\cap\mathcal U|>\sigma_i$.
    \item If no token conflicts with $\mathcal U$, the solver closes $\mathcal U$, and $M$ decreases
          by one.
    \item Otherwise, every busiest rank whose proposed expert is a candidate of a conflicting token
          sets $j_r\leftarrow j_r+1$, and the search returns to step 2. The repair fails after eight
          attempts or when some $j_r$ exceeds $m_r$.
\end{enumerate}
These ranks advance together, so each attempt is one parallel step. The joint advance can skip a
feasible proposal that advancing one rank at a time would find.

\subsection{Solver Loop and Depth Cap}
\label{app:decode_loop}

The depth cap $D$ bounds the total reduction of $M$ within one MoE layer of one decode step. The
default $D=\infty$ imposes no cap. Let $\delta$ denote the reduction reached so far, starting from
$\delta=0$. In each round with $\Delta^\star>0$, the solver closes $\mathcal R(\Delta)$ with
\begin{equation}
\label{eq:decode_depth_limit}
\Delta=\min(\Delta^\star,\,D-\delta)
\end{equation}
and sets $\delta\leftarrow\delta+\Delta$. A round with $\Delta^\star=0$ runs the repair step
instead, and a successful repair sets $\delta\leftarrow\delta+1$. After each round, the solver
recomputes the slack, the active counts, $M$, and the closure costs. It stops when $\delta=D$ or
when the repair fails. Every round except the last lowers $M$ by at least one, so the loop ends
after at most $M_0+1$ rounds, where $M_0$ is the initial value of $M$. Once $M$ stops decreasing,
the solver closes no further experts, because such closures would move routing weight without
lowering $M$.

\section{Implementation Details}
\label{app:implementation}

Both phases predict the routing of the next MoE layer from the hidden states of the current one.
The first MoE layer has no predecessor, so its prefill price is the load-feedback offset alone and
its decode routing stays native Top-$k$. The last MoE layer computes no prediction. Only the two
online corrections carry state across forwards.

\subsection{Prefill Pipeline}
\label{app:prefill_pipeline}

The statistics kernel runs on the histogram stream of Figure~\ref{fig:rankcast_system}(a) and
reads a copy of $\mathbf X^{(\ell)}$. The main stream waits for it just before
\texttt{notify\_dispatch} reads the statistic row. \texttt{notify\_dispatch} already copies each
rank's token-count row to every other rank, and RankCast appends its statistic row to this copy. A
row holds $P(B+1)+1$ int32 values, or 8,240 bytes for $P=8$ and $B=256$ after padding to 16 bytes.
Across nodes, \texttt{notify\_dispatch} sends the row over the RDMA queue pair of the counts,
forwards it to the local ranks over NVLink, and sums the rows before it returns. A rank without
tokens sends an all-zero row, because every rank must join the exchange.

Once dispatch completes, the price kernel computes the target loads and raw prices of
Appendices~\ref{app:capacity} and~\ref{app:histogram_error} from the summed rows. It then subtracts
the mean price and adds the load-feedback offset of Appendix~\ref{app:online_corrections}. Integer
sums and a fixed order of floating-point operations make the prices bit-identical on every rank.
At the target layer, a fused Triton kernel replaces the native Top-$k$ and applies
Equation~\ref{eq:prefill_selection} with these prices.

\subsection{Online Corrections}
\label{app:online_corrections}

\paragraph{Score bias.}
The bias $\mathbf b^{(\ell+1)}$ of Equation~\ref{eq:cross_layer_prediction} tracks the mean gap
$\bar g_e$ between the target layer's router score of expert $e$ and its prediction without the
bias. Every 32 forwards, an all-reduce sums the gaps across ranks, and each rank updates
\begin{equation}
\label{eq:app_score_bias}
b_e\leftarrow(1-\nu)\,b_e+\nu\,\bar g_e,\qquad\nu=0.25.
\end{equation}

\paragraph{Load feedback.}
In prefill, a per-rank offset $\eta_q$ raises the price of ranks that stay overloaded. Every $T=8$
forwards, an all-reduce gives the number of expert assignments $n_q$ that each rank received in
recent forwards. Let $\phi_q=\log\bigl(P\,n_q/\sum_{q'}n_{q'}\bigr)$ denote the log load share of
rank $q$, which is zero under an even split, and let $\bar\phi$ denote its mean over ranks. The
offset moves toward balance:
\begin{equation}
\label{eq:app_load_feedback}
\eta_q\leftarrow\eta_q+\kappa\,(\phi_q-\bar\phi),\qquad\kappa=0.1.
\end{equation}
Each step is limited to $\pm0.069$, each offset to $\pm1$, and the offsets are kept centered. The
price kernel adds $\eta_q$ to the centered price $\beta_q$. Both corrections update outside CUDA
graphs at the end of a forward.

\subsection{Decode Candidate Rows}
\label{app:decode_payload}

Let $G$ denote the maximum number of decode tokens per rank. A row lists a token's Top-$L$
predicted experts in descending order, together with the 2-bit code $3\omega_{i,e}$ of each
(Appendix~\ref{app:decode_prefix_details}). The first $p$ entries form $\mathcal F_i$, and the
first $k$ form the initial selection. The rows travel in the token-count messages of DeepEP's
low-latency dispatch as two byte arrays:
\begin{equation}
\label{eq:decode_payload_layout}
\underbrace{[\mathrm{identifiers}]}_{PGL}
\ \Vert\
\underbrace{[\mathrm{codes}]}_{PG\lceil L/4\rceil\ \mathrm{bytes}}.
\end{equation}
For Qwen3-235B ($P=8$, $G=32$, $L=16$), they take 5~KiB per MoE layer. Each rank writes its rows
into its own slice of a zeroed buffer, and each receiver sums the buffers of all ranks. The slices
are disjoint, so the sum recovers every rank's rows. Across nodes, a rank writes its rows over the
RDMA queue pair of its token count before it sends the count. The queue pair delivers in order, so
the rows arrive before the count and need no separate signal.

\subsection{Decode Solver Kernel}
\label{app:decode_cuda_layout}

Each rank runs the solver as one CUDA block. The block keeps the candidate graph in shared memory
as bit matrices:
\begin{equation}
\label{eq:decode_bit_matrices}
\operatorname{cmask}[i]=\{e\mid e\in\mathcal{C}_i\},
\qquad
\operatorname{tokmask}[e]=\{i\mid e\in\mathcal{C}_i\}.
\end{equation}
Two more matrices, $\operatorname{mhi}[e]$ and $\operatorname{mlo}[e]$, mark the tokens whose code
for expert $e$ has its high or low bit set. Let $\operatorname{sbuck}[j]$ denote the tokens with
slack $j$, and let $\operatorname{popc}$ count set bits. The closure cost of
Equation~\ref{eq:decode_close_cost} then needs only population counts:
\begin{equation}
\label{eq:decode_bucketed_cost}
\operatorname{cost}(e)
=
\frac{1}{3}\sum_{j=0}^{L-k}
\frac{
2\operatorname{popc}(\operatorname{mhi}[e]\cap\operatorname{sbuck}[j])
+\operatorname{popc}(\operatorname{mlo}[e]\cap\operatorname{sbuck}[j])
}{j+1}.
\end{equation}
Scaling every cost by $3\operatorname{lcm}(1,\ldots,L-k+1)$ makes it an integer, so every rank
orders the experts identically. For each level, the solver keeps a mask of the experts at that
level. Each token counts its candidates level by level to find $\Lambda_i$, and a block-wide
minimum gives $\Delta^\star$. After the last round, the solver writes a router bias of $-10^4$ for
every closed expert and $0$ otherwise. The kernel supports up to 16 ranks, 512 candidate rows, and
32 experts per rank.

\paragraph{CUDA graphs and validation.}\label{app:decode_abi_validation}
The solver uses only preallocated buffers on the caller's stream and never synchronizes with the
host, so the decode path can be captured in a CUDA graph. It never stops on invalid input, because
a stopped rank would stall the dispatch of every other rank. A row with fewer than $k$ valid
candidates instead gets zero slack, which keeps that token's experts open. A scalar NumPy
reference reproduces the CUDA masks bit for bit on captured Qwen3-235B router scores and on
synthetic cases.

\section{Evaluation Details}
\label{app:eval_details}

\subsection{Settings and Protocols}
\label{app:eval_protocol}

\paragraph{System settings.} On Qwen3-235B, SGLang's mini-lb routes requests to the prefill and
decode workers, and Mooncake~\citep{qin2025mooncake} transfers the KV cache between them. Prefill
uses DeepEP normal mode without CUDA graphs, and decode uses DeepEP low-latency mode with CUDA
graphs. No run uses the radix cache. Online EPLB rebalances every 50 forward passes, and the other
configurations keep a static expert placement. RankCast changes only which experts a token uses,
and all of its communication stays inside the EP group. It neither adds traffic to the KV path nor
synchronizes with it.

\paragraph{Throughput runs.} Each run discards its warmup requests and then measures 300 seconds,
or 180--240 seconds for the natural-length decode workloads. Prefill baseline and RankCast runs
alternate, with two or three runs per configuration. The LongBench v2 and EPLB results are single
runs. Each decode configuration runs in one server launch. On Qwen3-235B, each natural-length
decode cell averages two runs, and the second run uses a new seed and the reverse order. All other
decode cells are single runs.

\paragraph{Prefill.} LongBench v2 inputs come from all six domains, with at most four 32,768-token
windows per document. Each request generates one output token, and the client keeps a fixed number
of requests in flight. On Qwen3-235B, this number is 64 for the 4,096-token workloads and 24 for
LongBench v2, which runs with TP attention. On Qwen3.5-397B, it is 128, 192, and 320 at the three
chunk sizes, and 32 for LongBench v2. The max-to-mean token ratio is averaged over all MoE layers
and over the forward passes that carry at least 90\% of the tokens of the largest one. The
small-chunk ShareGPT baseline on Qwen3-235B comes from other sessions, whose baseline throughput
varies by 0.4\%. Table~\ref{tab:prefill_abs} gives the absolute input throughput.

\begin{table}[t]
    \centering
    \caption{Prefill input throughput (thousand tokens/s) behind the gains of Table~\ref{tab:prefill}. $^\dagger$Run with a larger state cache for the linear-attention layers; the gain in Table~\ref{tab:prefill} is relative to a baseline with the same cache.}
    \label{tab:prefill_abs}
    \small
    \begin{tabular}{llcccc}
        \toprule
        Workload     & Tokens per rank & Baseline               & EPLB                    & EPLB-R                  & RankCast \\
        \midrule
        \multicolumn{6}{l}{\textit{Qwen3-235B}}                                                                                    \\
        ShareGPT     & 2K              & 44.6        & 50.3                    & 47.5                    & 51.9     \\
                     & 4K              & 48.7                   & 55.9                    & 52.6                    & 57.7     \\
                     & 8K              & 50.0                   & 57.0                    & 53.4                    & 59.7     \\
        \addlinespace[2pt]
        CodeContests & 2K              & 45.4                   & 51.2                    & 48.3                    & 52.8     \\
                     & 4K              & 50.1                   & 57.5                    & 53.6                    & 59.5     \\
                     & 8K              & 51.5                   & 58.8                    & 55.3                    & 61.4     \\
        \addlinespace[2pt]
        LongBench    & 2K              & 30.3                   & 30.2                    & 30.0                    & 32.5     \\
                     & 4K              & 31.8                   & 31.8                    & 31.7                    & 34.0     \\
                     & 8K              & 33.1                   & 33.4                    & 33.2                    & 34.5     \\
        \midrule
        \multicolumn{6}{l}{\textit{Qwen3.5-397B}}                                                                                  \\
        ShareGPT     & 4K              & 120.1                  & 125.8                   & 125.5                   & 133.1    \\
                     & 8K              & 123.2                  & 132.7                   & 132.4                   & 136.9    \\
                     & 16K             & 122.6                  & 136.1\rlap{$^\dagger$}  & 135.6\rlap{$^\dagger$} & 135.4    \\
        \addlinespace[2pt]
        CodeContests & 4K              & 120.5                  & 129.1                   & 128.3                   & 137.2    \\
                     & 8K              & 124.6                  & 137.0                   & 136.0                   & 142.8    \\
                     & 16K             & 125.1                  & 140.8\rlap{$^\dagger$}  & 139.9\rlap{$^\dagger$} & 142.6    \\
        \addlinespace[2pt]
        LongBench    & 8K              & 122.3                  & 122.0                   & 122.5                   & 127.4    \\
                     & 16K             & 121.4                  & 122.3                   & ---                     & 126.7    \\
        \bottomrule
    \end{tabular}
\end{table}

\begin{figure}[t]
    \centering
    \includegraphics[width=\linewidth]{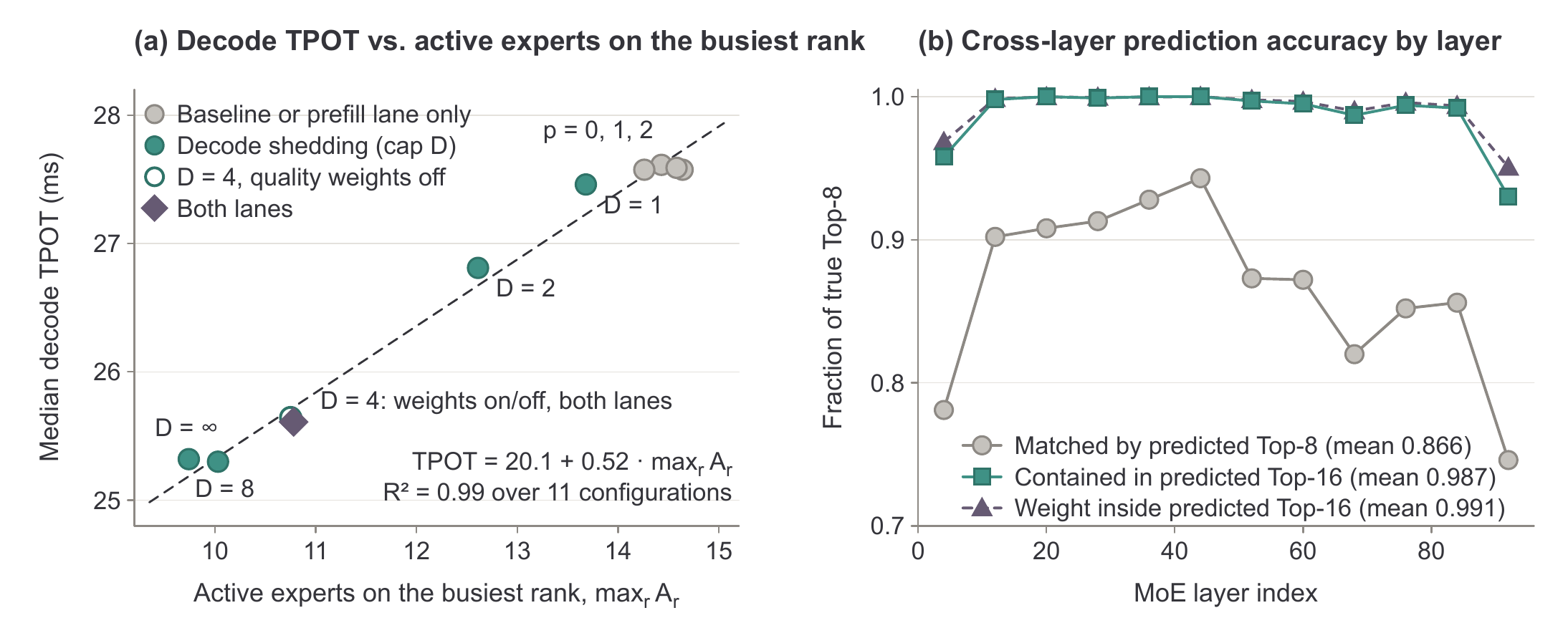}
    \caption{(a) Median decode TPOT against the mean number of active experts on the busiest rank for 11 configurations of Qwen3-235B at ISL 1024 and global batch 96. The dashed line is a least-squares fit. (b) Cross-layer prediction accuracy on 12 sampled MoE layers of Qwen3-235B during prefill. The curves give the share of each token's true Top-8 experts inside the predicted Top-8 and inside the predicted Top-16, and the share of their routing weight inside the predicted Top-16.}
    \label{fig:eval_decode_pred}
\end{figure}

\paragraph{Decode.} Each request generates 1,024 output tokens, with end-of-sequence tokens
ignored. In the tiled ShareGPT workload, requests arrive at a fixed rate, and the global batch size
caps the number in flight. The natural-length workloads keep the global batch size of requests in
flight, which is 96 on Qwen3-235B and 128 on Qwen3.5-397B. We report median TPOT, because output
throughput does not follow decode speed in these runs. The tiled workload nearly keeps up with its
arrival rate, and on short natural-length prompts the bottleneck lies outside decode. On
Qwen3.5-397B, TPOT counts only pure decode steps, which excludes the interleaved prefill chunks. The
active-expert count on the busiest rank is averaged over layers and steady-state decode steps. Its
run-to-run spread is about 0.2 experts, measured on the prefill-balancing runs, which do not change
decode routing. Figure~\ref{fig:eval_decode_pred}(a) and Table~\ref{tab:decode_ablation_quality}
come from an earlier set of runs on the tiled workload. In those runs, decode shedding without a
depth cap gained 9.1\%.

\begin{table}[t]
    \centering
    \caption{Offline comparison of decode shedding with XShare~\citep{vankov2026xshare} and OEA~\citep{oncescu2025opportunistic} on decode batches of Qwen3-235B. $M$ is the mean active-expert count on the busiest rank. Each row compares the methods at the same $M$. Every method keeps the true top-1 expert of at least 99\% of tokens and gives every token $k$ experts. The RankCast column gives the routing probability displaced from each token's Top-$k$. The XShare and OEA columns give their displaced probability as a multiple of RankCast's.}
    \label{tab:decode_baselines}
    \small
    \begin{tabular}{lcccc}
        \toprule
        Workload      & $M$   & RankCast   & XShare        & OEA           \\
        \midrule
        ShareGPT      & 14.56 & $0.065\%$  & $2.38\times$  & $1.94\times$  \\
                      & 13.59 & $0.180\%$  & $6.02\times$  & $4.75\times$  \\
                      & 12.01 & $0.600\%$  & $3.60\times$  & $3.83\times$  \\
                      & 11.59 & $0.811\%$  & $2.83\times$  & $2.84\times$  \\
        \addlinespace[2pt]
        LiveCodeBench & 12.96 & $0.062\%$  & $4.21\times$  & $1.92\times$  \\
                      & 12.05 & $0.133\%$  & $1.96\times$  & $2.25\times$  \\
                      & 10.25 & $0.428\%$  & $3.00\times$  & $4.15\times$  \\
                      & 9.22  & $0.731\%$  & $2.34\times$  & $2.43\times$  \\
        \bottomrule
    \end{tabular}
\end{table}

\paragraph{Offline decode comparison.} Table~\ref{tab:decode_baselines} replays router scores
captured from Qwen3-235B decode. Each workload uses two disjoint batches of 96 sequences, with 64
decode steps over all 94 MoE layers. The ShareGPT capture uses 1,024-token contexts, and the
LiveCodeBench capture uses 2,048-token contexts. Native routing gives $M=15.61$ on ShareGPT and
$M=14.11$ on LiveCodeBench. Every method closes experts with the additive bias
of Section~\ref{sec:decode_implementation}. The native Top-$k$ then runs on the biased true scores.
RankCast uses $p=2$, $L=16$, and its cross-layer prediction, and its rows come from different depth
caps $D$. RankCast protects the predicted top experts, so it keeps the true top-1 expert of
99.2--99.9\% of tokens. XShare follows Algorithm~6 of its paper, and OEA follows its published
algorithm. Both
receive the true router scores of the batch. A baseline matches a RankCast row when its $M$ is at
most 0.15 above RankCast's.

\paragraph{Quality.} The quality suite comprises MMLU-Redux 2.0~\citep{gema2025mmlu},
MATH-500~\citep{hendrycks2021math,lightman2023verify}, MMMLU~\citep{openai2024mmmlu},
RULER~\citep{hsieh2024ruler}, GPQA-Diamond~\citep{rein2024gpqa}, AIME 2025~\citep{maa2025aime},
IFEval strict~\citep{zhou2023ifeval}, and LiveCodeBench
v5~\citep{jain2024livecodebench,livecodebench2025releasev5}. MMMLU uses a fixed 10\% stratified
subset of each of its 14 languages. RULER uses 100 instances of each of its 13 tasks at 4K, 16K,
and 32K tokens. Quality runs use 2K tokens per rank on Qwen3-235B and 1K on Qwen3.5-397B, with
server seed 20260809. Generation is non-thinking and samples with temperature 0.7 and nucleus
probability 0.8 from the 20 most likely tokens. RULER uses greedy decoding. Each repeat starts a
new server session. On Qwen3-235B, MMLU-Redux 2.0, MATH-500, MMMLU, and RULER use three repeats,
and the other four benchmarks use ten. Decode shedding uses 13 repeats on GPQA-Diamond and AIME
2025. On Qwen3.5-397B, prefill balancing uses the same counts, except five repeats on AIME 2025.
Decode shedding uses two repeats on the first four benchmarks, ten on AIME 2025, and five on the
other three. Qwen3.5-397B runs both phases in one instance, so it enables only one of the two at a
time. The change of repeat $r$ is the score of repeat $r$ minus the baseline score of repeat $r$.
Its standard deviation is 0.5--1.3\,pp on the four ten-repeat benchmarks, which are also the four
smallest. On MMLU-Redux 2.0, MMMLU and MATH-500 it is at most 0.23\,pp.

\begin{table}[t]
    \centering
    \caption{Largest quality drop of each benchmark over all settings of Table~\ref{tab:quality}, against the baseline spread, the 95th percentile of the difference between two halves of the ten baseline repeats over all splits. The four benchmarks with three repeats have no such split. Percentage points.}
    \label{tab:quality_noise}
    \small
    \setlength{\tabcolsep}{5pt}
    \begin{tabular}{lrcccc}
        \toprule
                         &           & \multicolumn{2}{c}{Qwen3-235B} & \multicolumn{2}{c}{Qwen3.5-397B} \\
        \cmidrule(lr){3-4}\cmidrule(lr){5-6}
        Benchmark        & Questions & Largest drop & Baseline spread & Largest drop & Baseline spread \\
        \midrule
        GPQA-Diamond     & 198       & 0.85 & 1.35 & 0.47 & 1.20 \\
        AIME 2025        & 30        & 0.54 & 0.89 & 0.09 & 1.57 \\
        IFEval strict    & 541       & 0.65 & 0.65 & 0.00 & 0.67 \\
        LiveCodeBench v5 & 880       & 0.84 & 0.89 & 0.43 & 0.66 \\
        \addlinespace[2pt]
        MMLU-Redux 2.0   & 5,700     & 0.21 & ---  & 0.01 & --- \\
        MATH-500         & 500       & 0.40 & ---  & 0.30 & --- \\
        MMMLU            & 19,600    & 0.22 & ---  & 0.25 & --- \\
        RULER            & 3,900     & 0.73 & ---  & 0.20 & --- \\
        \bottomrule
    \end{tabular}
\end{table}

\paragraph{Measurement resolution.}\label{sec:eval_stats} On MMLU-Redux 2.0 and MMMLU the 95\%
confidence interval of every change stays within $\pm0.5$\,pp, so three repeats resolve changes of
this size. The four ten-repeat benchmarks vary more. Table~\ref{tab:quality_noise} therefore reports
a baseline spread for each of them, the largest gap the benchmark shows when nothing has changed. We
split the ten baseline repeats into two halves, take the difference between the halves, and report
its 95th percentile over all splits. The largest drop anywhere is 0.85\,pp, and on each small
benchmark the largest drop is at or below the spread. A paired test over per-question scores finds 3
of the 38 comparisons significant at the 0.05 level.

\paragraph{Prediction accuracy protocol.} Figure~\ref{fig:eval_decode_pred}(b) uses Qwen3-235B
prefill with 2K tokens per rank. It samples the 12 target layers $4,12,\ldots,92$ over two forward
passes on all eight ranks, 393,216 token--layer pairs in total. For token $i$, let $\hat s_i$ denote
its predicted scores and $s_i$ its router scores at the target layer. The Top-$L$ recall is
\begin{equation}
\label{eq:app_topl_recall}
\frac{\bigl|\operatorname{Top}_8(s_i)\cap\operatorname{Top}_L(\hat s_i)\bigr|}{8},
\end{equation}
and the weight recall is the share of the routing weight of $\operatorname{Top}_8(s_i)$ that falls
on experts inside $\operatorname{Top}_L(\hat s_i)$.

\subsection{GEMM Profiling}\label{sec:eval_profile}

\begin{figure}[t]
    \centering
    \includegraphics[width=\linewidth]{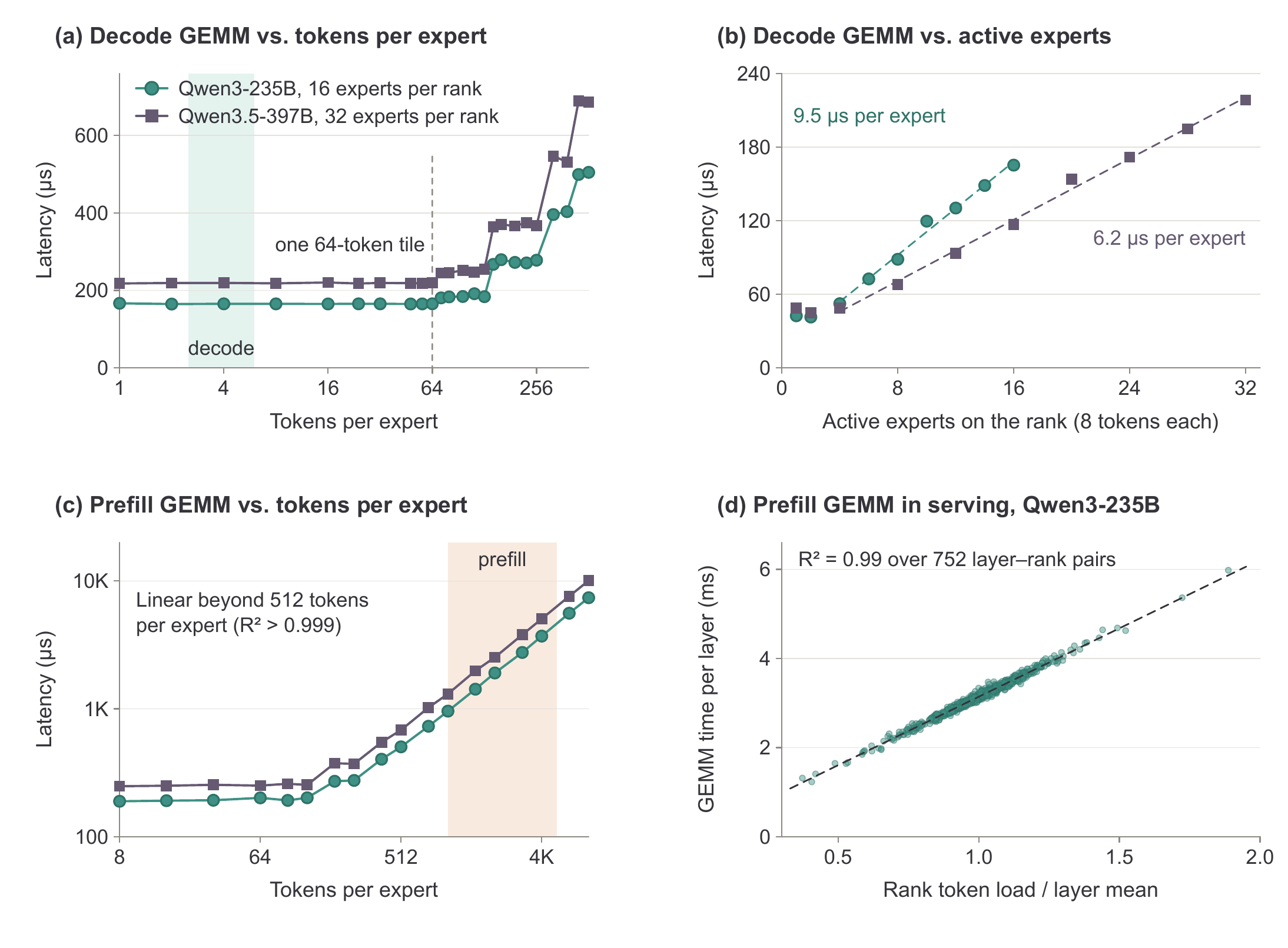}
    \caption{GEMM profiling. (a--c) Offline latency of the grouped expert GEMM for one MoE layer on one GPU, with the expert shapes and experts per rank of Qwen3-235B at EP$=8$ and Qwen3.5-397B at EP$=16$. (a) Decode layout with every expert on the rank active. (b) Decode layout with 8 tokens per active expert. (c) Prefill layout with every expert on the rank active. The shaded bands mark the tokens per expert at the batch sizes of Table~\ref{tab:decode_throughput} and at the chunk sizes of Table~\ref{tab:prefill}. (d) Expert GEMM time of each MoE layer and EP rank during Qwen3-235B prefill at 8K tokens per rank, against the token load of the rank divided by the layer mean. The dashed lines are least-squares fits.}
    \label{fig:gemm_profile}
\end{figure}

We profile the grouped expert GEMM of DeepGEMM, which computes both expert projections of one MoE
layer on one rank. Figure~\ref{fig:gemm_profile}(a--c) measures it offline on one GPU with the
expert shapes and experts per rank of both models. Decode and prefill lay out the tokens
differently, so we measure each layout.

Figure~\ref{fig:gemm_profile}(a) varies the tokens per expert in the decode layout while every
expert on the rank is active. Latency changes by less than 1.3\% from 1 to 64 tokens per expert,
because one 64-token tile holds all tokens of an expert and the time goes to reading its weights.
The decode batches of Table~\ref{tab:decode_throughput} place 2.5--6 tokens on an average expert.
Decode GEMM time is therefore set by the number of active experts.
Figure~\ref{fig:gemm_profile}(b) shows that each active expert adds 9.5\,$\mu$s on Qwen3-235B and
6.2\,$\mu$s on Qwen3.5-397B. Their ratio of about 1.5 matches the ratio of their expert
intermediate sizes, 1,536 and 1,024.

Figure~\ref{fig:gemm_profile}(c) repeats the token sweep in the prefill layout. Beyond 512 tokens
per expert, latency grows linearly with the token count ($R^2>0.999$). The prefill chunks of
Table~\ref{tab:prefill} place 1,024--5,120 tokens on an average expert. Prefill GEMM time is
therefore set by the token load of the rank. Figure~\ref{fig:gemm_profile}(d) confirms this in
serving on Qwen3-235B. The GEMM time of each layer and rank grows linearly with the token load of
the rank ($R^2=0.99$). At the mean load, the part that does not depend on the load is 2.2\% of the
GEMM time.

\subsection{Ablations}
\label{app:eval_extra}

\paragraph{Protected prefix $p$.} The protected prefix costs no prefill throughput. We vary $p$ on
Qwen3-235B with ShareGPT 4K inputs at 2K tokens per rank. Input throughput differs by at most
0.3\% across $p=0$, $1$, and $2$. Imbalance differs by less than 0.01. Decode is more sensitive,
because the mask is applied before the target layer runs. The largest benchmark loss is 1.32\,pp at
$p=1$ and 0.45\,pp at $p=2$, so decode uses $p=2$ on both models.

\paragraph{Depth cap $D$ and quality weights.} Table~\ref{tab:decode_ablation_quality} gives the
throughput and quality of decode shedding on Qwen3-235B for each depth cap $D$ with quality
weights, and for $D=4$ without them. Caps of $D\leq4$ give a smaller decode gain than no cap. A cap
of $D=8$ gives the same gain as no cap. LiveCodeBench accuracy shows no monotone trend in $D$. A
depth cap therefore brings no benefit, and decode runs without one. At $D=4$, setting every
quality weight to one leaves the active-expert count on the busiest rank unchanged. LiveCodeBench
then drops by 1.49\,pp, against 0.45\,pp with the quality weights. The quality
weights therefore change which experts close, not how many.

\begin{table}[t]
    \centering
    \caption{Decode shedding on Qwen3-235B for each depth cap $D$ with quality weights, and for $D=4$ without them, at ISL 1,024 and global batch 96. $\max_r A_r$ is the mean number of active experts on the busiest rank. This count and the gain in $1/\mathrm{TPOT}$ come from the throughput runs. Quality changes are in percentage points. LiveCodeBench and IFEval use ten repeats, or 13 for $D=4$ with quality weights. MMLU-Redux uses three.}
    \label{tab:decode_ablation_quality}
    \small
    \begin{tabular}{llccccc}
        \toprule
        $D$      & Weights & $\max_r A_r$ & $\Delta\,1/\mathrm{TPOT}$ & LiveCodeBench v5 & IFEval strict & MMLU-Redux 2.0 \\
        \midrule
        1        & on      & 13.68        & $+0.6\%$                  & $-1.00$          & $-0.35$       & $-0.23$        \\
        2        & on      & 12.61        & $+3.0\%$                  & $-0.31$          & $+0.04$       & $-0.20$        \\
        4        & on      & 10.76        & $+7.7\%$                  & $-0.45$          & $-0.11$       & $-0.21$        \\
        8        & on      & 10.03        & $+9.1\%$                  & $-1.27$          & $-0.13$       & $-0.10$        \\
        $\infty$ & on      & 9.74         & $+9.1\%$                  & $-0.59$          & $-0.65$       & $-0.13$        \\
        4        & off     & 10.75        & $+7.7\%$                  & $-1.49$          & $-0.50$       & $-0.26$        \\
        \bottomrule
    \end{tabular}
\end{table}

\paragraph{Candidate width $L$.} Table~\ref{tab:candidate_width} recomputes the prediction accuracy
of Figure~\ref{fig:eval_decode_pred}(b) offline for each $L$. From $L=12$ to $L=32$, every four
added candidates roughly halve the missed routing weight. The decode candidate rows grow linearly
with $L$. The prefill statistic row does not depend on $L$. Raising $L$ from 16 to 24 enlarges the
candidate rows by 50\% and removes 73\% of the missed weight. Raising $L$ to 32 doubles the rows and
removes 90\%. $L=16$ already keeps 99.1\% of the routing weight, so Qwen3-235B uses $L=16$.

\begin{table}[t]
    \centering
    \caption{Cross-layer prediction accuracy on Qwen3-235B against the candidate width $L$, pooled over the 12 sampled layers of Figure~\ref{fig:eval_decode_pred}(b). Top-$L$ recall is the fraction of true Top-8 experts inside the predicted Top-$L$. Weight recall is the share of routing weight on these experts. Missed weight is the remainder. $L=16$ is the default on Qwen3-235B.}
    \label{tab:candidate_width}
    \small
    \begin{tabular}{lcccccc}
        \toprule
        $L$                & 8     & 12    & 16    & 20    & 24    & 32    \\
        \midrule
        Top-$L$ recall     & 0.866 & 0.966 & 0.987 & 0.994 & 0.997 & 0.999 \\
        Weight recall      & 0.903 & 0.976 & 0.991 & 0.996 & 0.998 & 0.999 \\
        Missed weight (\%) & 9.7   & 2.4   & 0.92  & 0.45  & 0.25  & 0.095 \\
        \bottomrule
    \end{tabular}
\end{table}

\paragraph{Histogram bins $B$ and the EP degree $P$.} Table~\ref{tab:ep_scaling} replays the prefill
prices offline on captured router scores, which extends the measured range of $P$. Native imbalance
grows quickly with $P$, because each rank holds fewer experts, and RankCast still removes about half
of the excess at $P=128$. The bin count hardly matters. Up to $P=128$, $B=64$ and $B=256$ give the
same imbalance to within 0.012 on Qwen3.5-397B. The exchanged statistic row stays small, 8~KiB at
$P=8$ with $B=256$ and 32~KiB at $P=128$ with $B=64$. It rides on the existing
\texttt{notify\_dispatch} message, so it adds no synchronization. Each rank computes its price from
the summed histogram in a fixed number of steps, with no iteration between ranks.

\begin{table}[htb]
    \centering
    \caption{Offline replay of the prefill prices at larger EP degrees. Imbalance is max-to-mean, native $\rightarrow$ RankCast, with the share of the excess over the mean removed in parentheses. Qwen3-235B uses 16,384-token prefill chunks; Qwen3.5-397B uses decode-shaped captures combined into 16,384-token batches.}
    \label{tab:ep_scaling}
    \small
    \begin{tabular}{lll}
        \toprule
        $P$ & Qwen3-235B, prefill chunk    & Qwen3.5-397B, code           \\
        \midrule
        8   & $1.60\rightarrow1.19$ (68\%) & ---                          \\
        16  & $2.30\rightarrow1.36$ (70\%) & $2.48\rightarrow1.39$ (72\%) \\
        32  & $3.45\rightarrow1.86$ (64\%) & $3.62\rightarrow1.87$ (66\%) \\
        64  & $5.30\rightarrow2.97$ (53\%) & $5.53\rightarrow2.92$ (57\%) \\
        128 & ---                          & $9.34\rightarrow5.30$ (51\%) \\
        \bottomrule
    \end{tabular}
\end{table}

\end{document}